\documentclass[arguments]{aastex631}

\usepackage{multirow}
\shorttitle{NGC7213}
\shortauthors{Shi et al.}
\begin{document}
%\title{Template \aastex Article with Examples: 
%v6.31\footnote{Released on March, 1st, 2021}}

%\title{A survey for low-luminosity Active Galactic Nuclei with hot wind in Chandra archive}
%\title{Searching for hot wind in low-luminosity active galactic nuclei}
%\title{X-ray spectroscopic study of hot wind in low-luminosity active galactic nucleus NGC 7213}
\title{Evidence for A Hot Wind from High-resolution X-ray Spectroscopic Observation of the Low-luminosity Active Galactic Nucleus in NGC 7213}

\author[0000-0003-3922-5007]{Fangzheng Shi}
\affiliation{School of Astronomy and Space Science, Nanjing University, Nanjing, China}
\affiliation{Key Laboratory of Modern Astronomy and Astrophysics, Nanjing University, Nanjing, China}
\email{fzshi@smail.nju.edu.cn}

\author[0000-0003-0900-4481]{Bocheng Zhu}
\affiliation{Shanghai Astronomical Observatory, Chinese Academy of Sciences, Shanghai, China}
\affiliation{School of Astronomy and Space Sciences, University of Chinese Academy of Sciences, Beijing, China}

\author[0000-0003-0355-6437]{Zhiyuan Li}
\affiliation{School of Astronomy and Space Science, Nanjing University, Nanjing, China}
\affiliation{Key Laboratory of Modern Astronomy and Astrophysics, Nanjing University, Nanjing, China}
\email{lizy@nju.edu.cn}

\author[0000-0003-3564-6437]{Feng Yuan}
\affiliation{Shanghai Astronomical Observatory, Chinese Academy of Sciences, Shanghai, China}
\affiliation{School of Astronomy and Space Sciences, University of Chinese Academy of Sciences, Beijing, China}
\email{fyuan@shao.ac.cn}

%\collaboration{6}{(AAS Journals Data Editors)}



\begin{abstract}
Super-massive black holes (SMBHs) spend most of their lifetime accreting at a rate well below the Eddington limit, manifesting themselves as low-luminosity active galactic nuclei (LLAGNs). The prevalence of a hot wind from LLAGNs is a generic prediction by theories and numerical simulations of black hole accretion and is recently becoming a crucial ingredient of AGN kinetic feedback in cosmological simulations of galaxy evolution. However, direct observational evidence for this hot wind is still scarce. In this work, we identify significant Fe XXVI Ly$\alpha$ and Fe XXV K$\alpha$ emission lines from high-resolution Chandra grating spectra of the LLAGN in NGC\,7213, a nearby Sa galaxy hosting a $\sim10^8\rm~M_\odot$ SMBH, confirming previous work. We find that these lines exhibit a blueshifted line-of-sight velocity of $\sim1100\rm~km s^{-1}$ and a high XXVI Ly$\alpha$ to XXV K$\alpha$ flux ratio implying for a $\sim16$ keV hot plasma. By confronting these spectral features with synthetic X-ray spectra based on our custom magnetohydrodynamical simulations, we find that the high-velocity, hot plasma is naturally explained by the putative hot wind driven by the hot accretion flow powering this LLAGN. Alternative plausible origins of this hot plasma, including stellar activities, AGN photoionization and the hot accretion flow itself, are quantitatively disfavored. The inferred kinetic energy and momentum carried by the wind can serve as strong feedback to the environment. We compare NGC\,7213 to M81*, in which strong evidence for a hot wind was recently presented, and discuss implications on the universality and detectability of hot winds from LLAGNs.
\end{abstract}



\section{Introduction} \label{sec:intro}
Super-massive black holes (SMBH) co-evolve with their host galaxies. This is evidenced by the strong correlations between the black hole mass and global properties (bulge mass, velocity dispersion and luminosity) of the host galaxy \citep{2009ApJ...698..198G,2011MNRAS.413.1479S,2013ARA&A..51..511K}.
Radiation, jets and outflows from an active galactic nucleus (AGN) can impose an enormous amount of energy into the environment of the SMBH on various physical scales, suppressing cooling flows and quenching star formation. Both theoretical and observational studies have suggested that AGN feedback is the primary driver for the co-evolution between SMBHs and their host galaxies (see reviews by \citealp{2012ARA&A..50..455F} and \citealp{2013ARA&A..51..511K}).
%Cosmological simulations of galaxy evolution on large scales implement the AGN feedback with sub-grid model that mimic the net effect on resolved scales \citep{2017MNRAS.465.3291W}. 
%The adopted sub-grid physics requires validation from observations probing into the origin of outflows in individual SMBH and the interaction between the outflows and circumnulear environment.

There exist two modes of black hole accretion, known as the cold mode and the hot mode \citep{2014ARA&A..52..529Y}, which correspond to two respective modes of AGN feedback: radiative mode and kinetic mode.
The radiative mode (or quasar mode) operates mainly in luminous AGNs whose  accretion rates are close to the Eddington limit. Radiatively driven ultra-fast outflows (UFOs) have been detected in the X-ray spectrum of up to 35\% of local luminous, radio-quiet AGNs, which are characterized by Fe XXV/XXVI absorption lines with blueshifted line-of-sight velocities of $\gtrsim10^4\rm~km~s^{-1}$ \citep{2010A&A...521A..57T}. 
However, in the local universe luminous AGNs are found in $\lesssim10\%$ of normal galaxies \citep{2013ApJ...768....1M},
while the majority of present-day SMBHs grow their mass at low rates ($\lesssim$1\% Eddington limit) via a radiatively inefficient, hot accretion flow \citep{2014ARA&A..52..529Y}, manifesting themselves as low-luminosity active galactic nuclei (LLAGNs; \citealp{2008ARA&A..46..475H}).
%Most galaxies are quiescent (like SgA$^*$) or radio-loud 
The kinetic mode (or jet mode) operates in these LLAGNs, which are typically radio-loud sources due to relativistic jets symbiotic with the hot accretion flow. 
%Similar to black hole binary, the inner radius of the supermassive black hole accretion thin disk would be truncated and replaced by radiatively inefficient hot accretion flow when the accretion rate is well below 2\% of the Eddington limit \citep{2014ARA&A..52..529Y}. 

Theories \citep{1999MNRAS.303L...1B} and numerical simulations \citep{2012MNRAS.426.3241N,2012ApJ...761..130Y,2012ApJ...761..129Y,2015ApJ...804..101Y,2021ApJ...914..131Y} matured over the past two decades predict that an energetic outflow in the form of an uncollimated hot wind must also be generated from the hot accretion flow. The wind has a mass outflow rate much larger than the accretion rate at the black hole horizon, thus playing a crucial role in the accretion process. Moreover, by depositing its energy and momentum to the environment, the hot wind can provide an efficient means of AGN kinetic feedback, in addition to the conventional jet-driven feedback. In fact, numerical simulations involving both the jet and the hot wind have shown that the momentum flux of the wind is always larger than that of jet \citep{2021ApJ...914..131Y}. 
The role of the hot wind in star formation and black hole growth has been investigated in detail by \citet{2019ApJ...885...16Y}.
In the influential cosmological simulations of galaxy formation and evolution, IllustrisTNG \citep{2017MNRAS.465.3291W, 2018MNRAS.473.4077P}, winds from weakly accreting SMBHs are invoked to quench star formation in intermediate- to high-mass galaxies.

However, direct observational evidence of the putative hot wind remains scarce.
Using {\it Chandra} non-dispersed spectrum, \cite{2013Sci...341..981W} showed that the accretion flow onto the Galactic center SMBH, Sgr A*, is consistent with an inward decreasing accretion rate, inferring the existence of an outflow, but no direct information about the outflow was yielded.
\citet{2016Natur.533..504C} inferred the existence of nuclear winds in a sample of quiescent galaxies hosting LLAGNs, based on optical spectroscopic observations that reveal kpc-scale bipolar emission features co-aligned with
strong velocity gradients of ionized gas.
Most recently, \citet{2021NatAs...5..928S} reported evidence for a high-velocity ($\sim$3000 km~s$^{-1}$), high-temperature ($\sim$ 10 keV) outflow from M81*, a nearby prototypical LLAGN, based on a pair of symmetrically blueshifted and redshifted Fe XXVI Ly$\alpha$ emission lines in the high-resolution {\it Chandra} grating spectrum of this bright X-ray source.
This outflow was interpreted as the long-sought hot wind produced in the hot accretion flow onto M81$^*$, which propagates out to $\gtrsim10^{6}$ times the gravitational radius of the SMBH.

The finding of \citet{2021NatAs...5..928S} has begun to bridge a gap between observations of LLAGNs and the successful theory of hot accretion flows, as well as modern cosmological simulations of galaxy formation that require a new kinetic feedback mode from weakly accreting SMBHs, i.e., in the form of a wind.
However, M81* remains the only LLAGN in which compelling evidence of a hot wind is found. 
In this work, we are thus motivated to search for signatures of the putative hot wind in more LLAGNs.
X-ray spectra of high spectral and angular resolutions afforded by {\it Chandra} have proved to be the key of revealing the hot wind. We have examined existing {\it Chandra} grating observations of local LLAGNs (see details in Appendix \ref{sec:sample}). Only one LLAGN, which resides in NGC\,7213, stands out as a promising candidate for a detailed analysis presented below. 

Lying at a distance of 22.75 Mpc ($z$=0.00584, from the NASA/IPAC Extragalactic Database\footnote{http://ned.ipac.caltech.edu}),
NGC 7213 is classified as a Sa galaxy with both Syfert and LINER (low-ionization nuclear emission line region) signatures, presumably powered by a central SMBH with a mass of $\sim10^8\rm~M_{\odot}$ estimated from stellar velocity dispersion \citep{2002ApJ...579..530W}.
Using optical integral-field spectroscopic observations of the inner galaxy, 
\cite{2014MNRAS.438.3322S} estimated a black hole mass of $8_{-6}^{+16}\times10^7\rm~M_{\odot}$ from the $M_{\rm BH}-\sigma$ relation of \cite{2009ApJ...698..198G}. Hereafter we adopt $M_{\rm BH}=10^8\rm~M_{\odot}$ as the fiducial black hole mass.
The bolometric luminosity of the LLAGN is estimated to be $9\times10^{42}\rm~erg~s^{-1}$ by \cite{2005MNRAS.356..727S} and $1.7\times10^{43}\rm~erg~s^{-1}$ by \cite{2012MNRAS.424.1327E} based on broadband spectral energy distribution fitting. These yield an average Eddington ratio of  $\lambda_{\rm Edd}=L_{\rm bol}/L_{\rm Edd}\sim10^{-3}$ ($L_{\rm Edd} = 1.3\times10^{38}[M_{\rm BH}/{\rm M_\odot}]\rm~erg~s^{-1}$ is the Eddington limit), which ensures the LLAGN classification. 

The LLAGN in NGC\,7213 has been extensively studied in the X-ray band. 
Spectra obtained by XMM-Newton and {\it Chandra} observations exhibit a number of prominent emission lines, in particular the K$\alpha$ line from neutral or weakly-ionized Fe and  highly-ionized iron lines including Fe XXV K$\alpha$ and Fe XXVI Ly$\alpha$ bwteen 6--7 keV \citep{2003A&A...407L..21B,2005MNRAS.356..727S,2008MNRAS.389L..52B,2013MNRAS.429.3439E}. 
In the {\it Chandra} high-resolution grating spectrum, the Fe XXVI Ly$\alpha$ and XXV K$\alpha$ lines appear blueshifted at a substantial velocity of $\sim900\rm~km~s^{-1}$ and were suggested to originate from a collisionally ionized gas \citep{2008MNRAS.389L..52B}. 
We shall revisit these lines in detail below to provide a physical understanding of their origin.
%The resonant Fe XXV K$\alpha$ lines was stronger than the forbidden line, indicating an origin of collisional ionization \citep{2013MNRAS.429.3439E}.
%The long-term X-ray light curve of the LLAGN in NGC 7213 exhibits a fast-rise-expotential-decay behaviro over the past $\sim38$ years \citep{2018MNRAS.475.1190Y}, which was interpreted as a long-lived, delayed tidal disruption event.
%The 2--10 keV X-ray emission of the LLAGN in NGC\,7213 follows bluer-when-brighter trend typical in LLAGNs \citep{2012MNRAS.424.1327E}.
%Strong Fe K$\alpha$ line with a width of FWHM$\sim2400\rm~km~s^{-1}$ has been detected in the X-ray spectrum with no Compton reflection component \citep{2008MNRAS.389L..52B,2015MNRAS.452.3266U}.
The underlying continuum shows no significant Compton reflection component \citep{2003A&A...407L..21B,2010MNRAS.408..551L,2015MNRAS.452.3266U}, as constrained by {\it BeppoSAX}, {\it Suzaku}, {\it Swift} and {\it NuSTAR} observations that cover the X-ray spectrum to above tens of keV. 
%\cite{2013MNRAS.429.3439E} deduced that the neutral Fe K$\alpha$ emission originates from Compton-thin material such as a broad-line region and 
Together with the narrow neutral Fe K$\alpha$ line, this suggests that 
the standard optically-thick accretion disk may be truncated at a radius of $10^3-10^4~r_{\rm g}$ \citep{2010MNRAS.408..551L}, where $r_{\rm g}=GM_{\rm BH}/c^2$ is the gravitational radius of the black hole. 
A hot accretion flow is likely responsible for the bulk of the X-ray emission from this LLAGN \citep{2010MNRAS.408..551L,2016MNRAS.463.2287X}, although contribution from a relativistic jet is also possible and indirectly suggested by the presence of a compact radio nuclear source \citep{2005MNRAS.356..727S}, which exhibits weakly correlated long-term variability with the X-ray nucleus \citep{2011MNRAS.411..402B}.
The four-decade-long X-ray light curve of the nucleus has also been associated with a fast-rise-exponential-decay behavior \citep{2018MNRAS.475.1190Y}, which was argued to be due to a delayed tidal disruption event.

The remainder of this paper is organized as follows. In Section \ref{sec:data}, we describe the relevant X-ray observations and our procedures of data reduction. Our X-ray spectral analysis, including a blind search of emission lines and characterization of the underlying thermal plasma, are presented in Section \ref{sec:result}. In Section \ref{sec:wind}, the observed X-ray spectrum exhibiting blueshifted lines from highly-ionized iron, is confronted with magnetohydrodynamical simulations to test the hot wind scenario. In Section \ref{sec:discuss},
we show that alternative origins of the observed Fe lines, such as stellar activity, AGN photoionization and hot accretion flow, are quantitatively disfavored, leaving the hot wind as the most plausible scenario. 
We also compare the cases of NGC\,7213 and M81* and address general implications for LLAGN feedback.
In Section \ref{sec:sum} we summarize our results.
Quoted errors throughout this work are at 90\% confidence level, unless otherwise stated. 
%Upper and lower limits are given at 3-$\sigma$ confidence level.

\section{Observations and data preparation} \label{sec:data}
In this work, we primarily utilized two sets of X-ray data of NGC\,7213: i) {\it Chandra} spectra obtained by the High Energy Transmission Grating (HETG) in combination with the Advanced CCD Imaging Spectrometer (ACIS), which are able to resolve emission lines expected from the hot plasma around the SMBH, at a velocity resolution of 300--2000 km~s$^{-1}$ over an energy range of 0.5--8 keV, and ii) {\it NuSTAR} spectra to constrain the broadband continuum over 3--79 keV.

\begin{figure}[htb!]
\centering\includegraphics[width=0.8\textwidth]{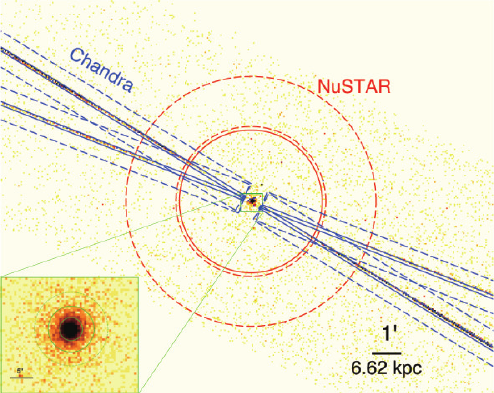}
\caption{A 0.5--8 keV ACIS/HETG image combining the two {\it Chandra} observations of NGC\,7213. 
The roll angles of the two observations differ by 180$^\circ$, hence the HEG/MEG dispersed arms of the two observations are indistinguishable here.  
The blue solid rectangles mark the spectral extraction region, with the adjacent dashed rectangles defining the background.  
The red solid and dashed circles mark the source and background spectral extraction regions for the {\it NuSTAR} observation.
The insert is a zoomed-in view of the circumnuclear region. The zeroth-order spectrum of the off-nucleus emission is extracted from the solid annulus, while the corresponding background is defined by the adjacent dashed annulus.
\label{fig:merge}}
\end{figure}

%\subsection{Chandra} \label{subsec:chandra}
NGC\,7213 was observed by {\it Chandra}/HETG on August 6 and 9, 2007 (ObsID 7742, 8590), with an exposure of 115 and 35 ks, respectively. 
The data was downloaded from the public archive and reprocessed following the standard procedure with CIAO v4.13 and calibration files CALDB v4.9.5.
The $\rm\pm1$-order grating spectra of both the high energy grating (HEG) and medium energy grating (MEG) were extracted using \emph{tgextract}, for each of the two observations.
For the source spectra,
we adopted the default cross-dispersion half-width of 2.4\arcsec (corresponding to 265 pc at the assumed distance of NGC\,7213), which ensures the same enclosed energy fraction ($\sim$97\%) for different wavelengths\footnote{https://cxc.cfa.harvard.edu/proposer/POG/}. 
The background spectra were extracted from two adjacent regions above and below the source region each with the default full-width of 19.1\arcsec. 
The spectral extraction regions are illustrated in Figure~\ref{fig:merge} displaying a 0.5--8 keV counts image combining the two ACIS/HETG observations.
Our visual examination of this image found no significant point sources that may contaminate the source and background spectra.  
The source shows no significant intra-observation and inter-observation variability. 
Hence we co-added the $\rm\pm1$-order spectra from the two observations, respectively for the HEG and MEG, which result in a total clean exposure time of 148 ks. 
The background spectra were also co-added, which contribute to $\lesssim2\%$ of the source counts.
We grouped the spectra to have at least one count per bin, to preserve the maximally possible spectral resolution for the subsequent spectral analysis.
%We note that the same spectra had been presented by \citet{2008MNRAS.389L..52B}, but only the portion of the spectrum containing the Fe lines (see below) were examined in that work.

%\subsection{NuSTAR} \label{subsec:nu}
NGC 7213 was targeted by {\it NuSTAR} on October 5, 2014, for an effective exposure of 101.6 ks (ObsID 6001031002). The data was downloaded and reprocessed following the standard \emph{nupipeline} in the software package NuSTARDAS v1.9.2. 
The spectra of NGC\,7213 were extracted for both focal plane modules A and B (FPMA and FPMB) with \emph{nuproducts}. The source region was a circle with a radius of 100\arcsec\ centered on the galaxy, while the background spectra were extracted from a concentric annulus with an inner radius of 105\arcsec\ and an outer radius of 175\arcsec\ (Figure~\ref{fig:merge}).
The spectra were similarly grouped to have at least one count per bin. 
Six off-nucleus point sources are present within the {\it NuSTAR} spectral extraction region. We find that their collective 0.5--8 keV flux accounts for less than 1\% of the flux of the nuclear source. Thus these off-nucleus sources should have a negligible contribution to the {\it NuSTAR} spectra.
We note that the source light curve shows no significant intra-observation variability.

\section{Spectral analysis} \label{sec:result}
\subsection{Broadband continuum} \label{subsec:cont}
\begin{figure}[htb!]
\centering\includegraphics[width=0.7\textwidth]{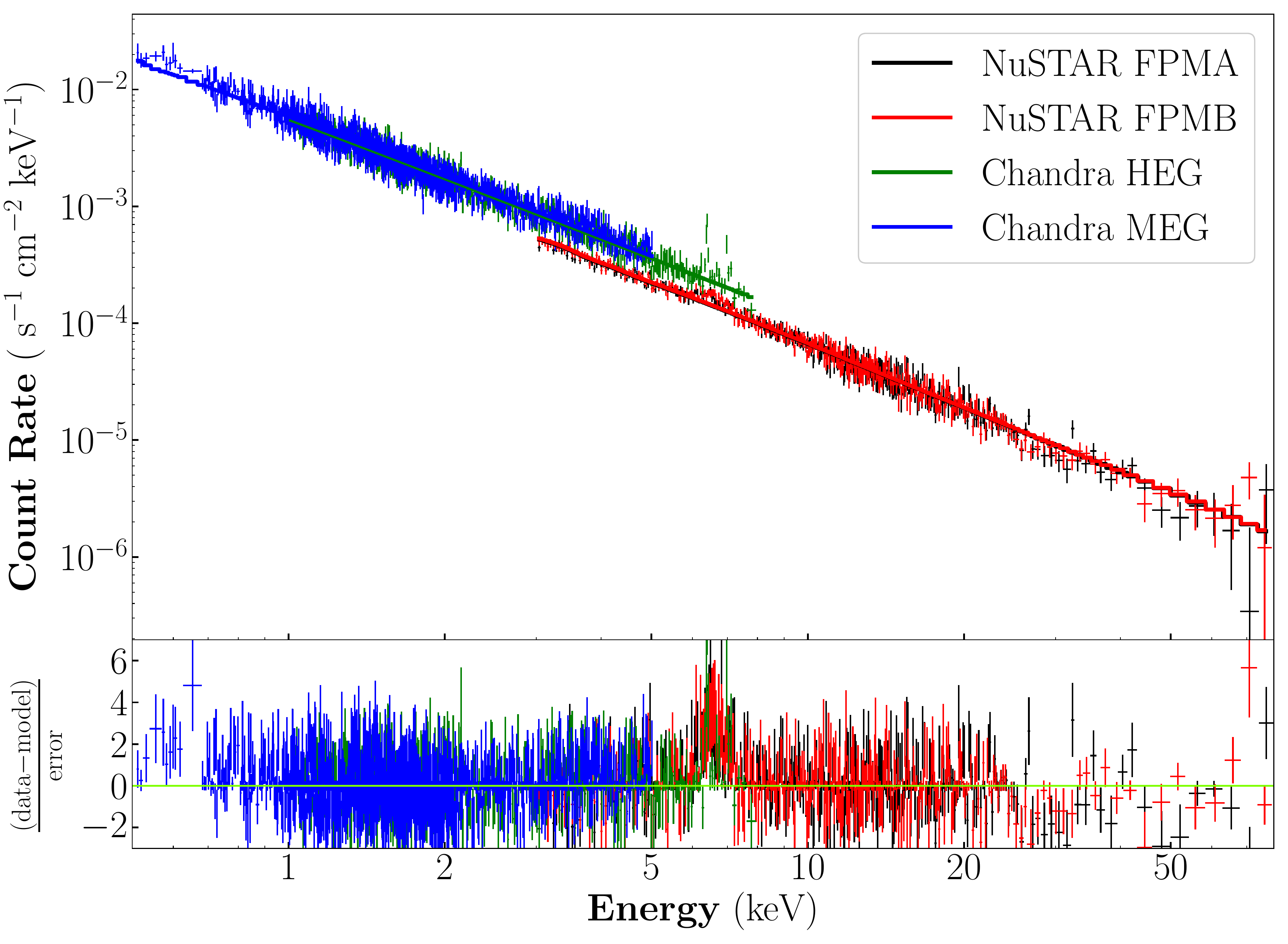}
\caption{Observed broadband X-ray spectra of NGC\,7213.
Blue: co-added {\it Chandra} MEG spectrum;
Green: co-added {\it Chandra} HEG spectrum;
Red: {\it NuSTAR} FPMA spectrum;
Black: {\it NuSTAR} FPMB spectrum.
Here the spectra are grouped to have a signal-to-noise ratio greater than 3 per bin for better visualization.
Error bars are of 1$\sigma$.
The spectra are fitted with an absorbed power-law. 
The lower panel shows the residual-to-error ratio. The energy range of 6--7.5 keV, where significant Fe lines are present, is excluded from the spectral fit.
\label{fig:baseline}}
\end{figure}

Spectral analysis is carried out with Xspec v.12.9.1, adopting the Cash statistics (\emph{C}-stat) to determine the best-fit models. 
To characterize the broadband X-ray continuum, we first consider a simple absorbed power-law model \emph{tbabs*powerlaw}, as we find no hint of excess foreground absorption in the spectra. 
%First, we checked the differences of spectra between two Chandra epochs. Their spectral indexes remain almost constant, indicating very little cross-observation spectral variabilities. Then, 
Hereafter the absorption column density is fixed at the Galactic foreground value, $N_{\rm H}=1.08\times10^{20}\rm~cm^{-2}$.
The co-added MEG spectrum over 0.5--5 keV, the co-added HEG spectrum over 1--8 keV,
and the FPMA/FPMB spectra over 3-79 keV are jointly fitted, excluding energies between 6--7.5 keV energy to avoid influence by the known Fe lines. 
%Since there was roughly an interval of 7 years between Chandra and NuSTAR observations, 
The photon-index $\Gamma$ is tied between HEG and MEG and between FPMA and FPMB, but allowed to be different between the {\it Chandra} and {\it NuSTAR} spectra.
The normalization is allowed to vary among the four spectra, to account for possible flux variation and difference in the enclosed-energy fraction.
The best-fit photon-index is $\Gamma=1.723\pm0.013$ for HEG/MEG and $\Gamma=1.838\pm0.012$ for FPMA/FPMB, which are consistent with previous studies \citep{2008MNRAS.389L..52B,2015MNRAS.452.3266U}. 
The point-spread function (PSF) corrected, unabsorbed 2-10 keV flux from the {\it Chandra} spectra is $\sim40\%$ higher than that from the {\it NuSTAR} spectra, in agreement with the flux decreasing trend found by \citet{2018MNRAS.475.1190Y}.  
The best-fit absorbed power-law models are plotted against the observed spectra in Figure \ref{fig:baseline}, and the fitted parameters are listed in Table \ref{tab:mod_fit}. 
%Table of apec fitting results
%[z-corrected]

%The flux and spectral variability follows the brighter-when-harder trend, as reported in \citep{2012MNRAS.424.1327E}, in opposite to the positive correlation between $\Gamma$ and $L_x$ observed mostly in luminous AGN. 
%However, the fluctuation of broadband X-ray flux for both NuSTAR and Chandra observations is less than 20\% on short timescale, as shown in the X-ray light-curves in Figure \ref{fig:lc}.

%\begin{figure}[hbt!]
%\plotone{lc_20211124.eps}
%\caption{Broadband X-ray light curve of NGC 7213 for NuSTAR observation (blue crosses) and Chandra observations (black crosses). The flux variability of NGC 7213 is less than 20\% with in 100-ks time-scale.
%\label{fig:lc}}
%\end{figure}

We then try to replace the power-law model with a cut-off power-law (\emph{cutoffpl} in Xspec), in order to test whether there is significant spectral softening at higher energies. We fit the {\it NuSTAR} spectra alone to constrain the high-energy cutoff, which results in a 3\,$\sigma$ lower limit, $E_{\rm cut}>93\rm~keV$, beyond the {\it NuSTAR} spectral coverage.
This suggests no significant spectral softening and that a single power-law is sufficient to account for the broadband continuum, especially at energies below 8 keV.
%{\bf By fixing the high-energy cutoff at this best-fit value, we also did a joint fit of HEG and MEG continuum with the cut-off power law model. The cut-off energy was fixed at the value derived in NuSTAR fitting. The best-fit photon index $\Gamma=1.713\pm0.013$ are consistent with the simple power law fitting result with no significant improvement in \emph{C}-stat.}
To further examine the possible presence of a reflection component, we employ the \emph{pexmon} model in Xspec, which can self-consistently calculate the reflected continuum from neutral gas and the fluorescent lines of neutral iron. 
In this model, the intrinsic continuum assumes a power-law with an exponential cutoff. 
We find a best-fit $\Gamma = 1.786^{+0.013}_{-0.005}$, $E_{\rm cut} = 184^{+32}_{-60}\rm~keV$,
%was consistent with the cut-off power law and simple power law fitting mentioned above. High-energy cut-off has also been constrained to $184^{+32}_{-60}\rm~keV$. 
%We found the reflection component contributes very little to the total continuum, with 
and a reflection fraction $R=\frac{\Omega}{2\pi}<0.06~(3\,\sigma)$. During the fit the inclination angle has been fixed at $45^{\circ}$ to reduce model degeneracy, and 
we have verified that changing this parameter has little effect on the derived reflection fraction. 
The lack of a reflection component, also found by \citet{2010MNRAS.408..551L} based on {\it Suzaku} observations, indicates a truncated accretion disk.
%Assuming a lamppost geometry, in which the illuminating source is located at $\sim100~r_{\rm g}$ above the accretion disk plane, the constraint on the reflection fraction indicates that the thin disk should be truncated at $\gtrsim1000~r_{\rm g}$.
%They suggested that the inner accretion disk may be truncated up to $10^{3}-10^{4}\rm~r_{g}$.

We conclude that the broadband X-ray spectra of the LLAGN in NGC\,7213 can be well characterized by a power-law model. 

\begin{deluxetable*}{lcccccccccc}
\footnotesize
%\tablenum{1}
\tablecaption{Broadband spectral fit results\label{tab:mod_fit}}
\tablewidth{0pt}
\tablehead{
\colhead{Data and model} & \colhead{$C$/d.o.f} & \colhead{$\Gamma$} & \colhead{$N_{\rm pl}$} & \colhead{$kT_{\rm h}$} & 
\colhead{$kT_{\rm l}$} &
\colhead{$v_{\rm h}$} &
\colhead{$v_{\rm l}$} &
\colhead{$N_{\rm h}$} &
\colhead{$N_{\rm l}$} & \colhead{$N_{\rm wind}$}
%\colhead{ } & \colhead{ } & \colhead{ } & \colhead{($10^{-3}\rm~ph~s^{-1}~cm^{-2}~keV^{-1}$)} & \colhead{(keV)} & \colhead{$\rm (km~s^{-1})$} & \colhead{$(10^{-3}\rm cm^{-5})$ & & &}
}
\colnumbers
\startdata
% TBABS*powerlaw
PL & & & & & & & & & & \\
\hline
HEG (1--8 keV) & 4367/ 4212 & \multirow{2}*{$1.723^{+0.013}_{-0.013}$} & $5.69^{+0.09}_{-0.09}$ & - & - & - & - & - & - & -\\
MEG (0.5--5 keV) & 3750/ 3881 & {                   } & $5.96^{+0.06}_{-0.06}$ & - & - & - & - & - & - & -\\
FPMA (3--79 keV) & 1665/ 1592 &  \multirow{2}*{$1.838^{+0.012}_{-0.012}$} & $4.76^{+0.12}_{-0.12}$ & - & - & - & - & - & - & -\\
FPMB (3--79 keV) & 1581/ 1570 & {            } & $4.80^{+0.12}_{-0.12}$ & - & - & - & - & - & - & -\\
\hline
% TBABS*(powerlaw+apec+zgauss+zgauss)
PL+apec$_{\rm h}$ & & & & & & & & & & \\
\hline
%HEG (1-8 keV) & 4283/ 4208 & 1.723 & 5.0^{+0.2}_{-0.3} & 13^{+3}_{-3} & 1.0^{+0.1}_{-0.5} & 2.2^{+0.9}_{-0.8} & - & - & -\\ 
HEG (5--8 keV) & 354/356 & 1.723 & $4.1^{+0.9}_{-1.1}$ & $16^{+6}_{-6}$ & - & $-1.1^{+0.3}_{-0.4}$ & - & $4.4^{+1.9}_{-1.8}$ & - & -  \\
\hline
PL+wind & & & & & & & & & &\\
\hline
HEG (5--8 keV) & 356/358 & 1.723 & $2.4_{-0.2}^{+0.2}$ & - & - & - & - & - & - & $0.2^{+0.1}_{-0.1}$ \\
\hline
PL+apec$_{\rm h}$+apec$_{\rm l}$ & & & & & & & & & & \\
\hline
HEG (1--8 keV) & 4287/4206 & \multirow{2}*{1.723} & \multirow{2}*{$5.1^{+0.2}_{-0.2}$}  & \multirow{2}*{$13^{+5}_{-2}$} & \multirow{2}*{$0.8^{+0.2}_{-0.1}$} &  \multirow{2}*{$-1.2^{+0.1}_{-0.2}$} & \multirow{2}*{$0.04^{+0.12}_{-0.03}$} & \multirow{2}*{$1.9^{+0.6}_{-0.7}$} & \multirow{2}*{$0.05^{+0.03}_{-0.01}$} & - \\
MEG (0.5--5 keV) & 3740/3875 & { } & { } & { } & { } & { } & { } & { } & { } & -  \\
\enddata
\tablecomments{(1) The fitted data set and the adopted model; `h' and `l' denote high- and low-temperature, respectively. `PL' stands for power-law and `wind' denotes a synthetic wind model; (2) Cash statistics over degree of freedom; (3) Power-law photon-index; (4) Normalization of the power-law, in units of $10^{-3}\rm~ph~s^{-1}~cm^{-2}~keV^{-1}$ at 1 keV; (5)-(6) Plasma temperature of high- and low-temperature \emph{apec} models; (7)-(8) Line-of-sight velocity of high- and low-temperature \emph{apec} model, in units of $10^3\rm~km~s^{-1}$; (9)-(10) Normalization of high- and low-temperature \emph{apec} model, in units of $10^{-3}\rm~cm^{-5}$; (11) Normalization of the synthetic wind model, in units of $10^{-3}\rm~cm^{-5}$.
% wind part (N_apec_l unit: 1e-3, N_wind unit:1e-4)
%$apec_{\rm h}$ refers to the hotter plasma with higher velocity and $apec_{\rm l}$ is the static plasma component with lower temperature.
Errors are quoted at 90\% confidence level. 
}
\end{deluxetable*}

%Weak evidence of jet structure in NGC 7213 has been found in 8.4 GHz radio observation(\citep{2005MNRAS.356..734B}). 
%The non-thermal X-ray enhancement could be explained by a jet toward our line-of-sight. However, the lack of X-ray variability within 100-ks time-scale disfavors the existence of jet.
%The power law component can also be contributed by the bremsstrahlung emission from the hot accretion inflows.
%In the following analysis searching for emission lines, we chose the simple power law to model the baseline continuum. 

\subsection{Blind search of lines} \label{subsec:blsearch}
\begin{figure}[hb!]
\centering\includegraphics[width=0.75\textwidth]{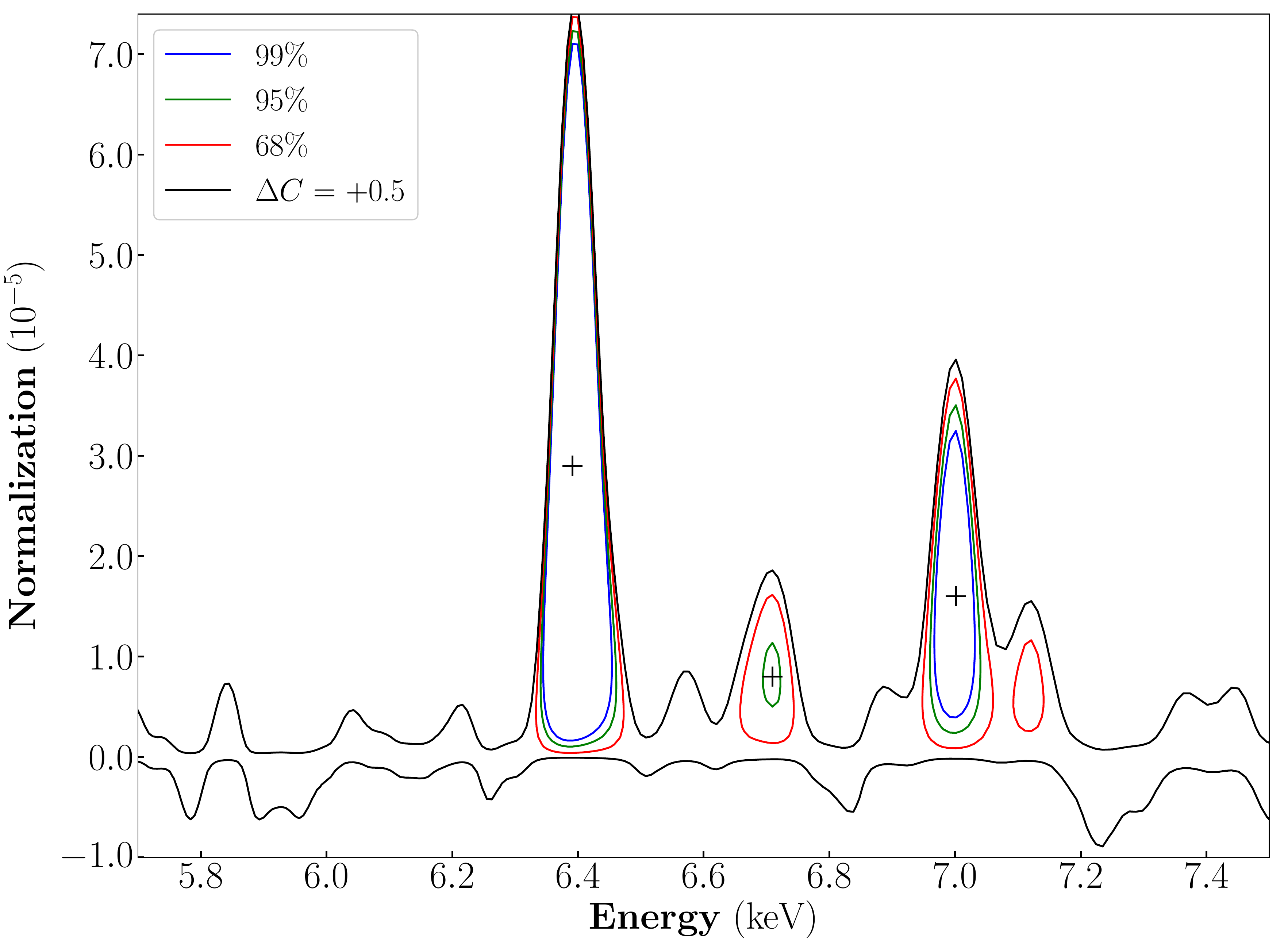}
\caption{Significant lines between 5--8 keV in the co-added HEG spectrum from a blind search.
The photon energy has been corrected for the systemic redshift of NGC\,7213.
Contours denote the improvement of $C$-stat ($\Delta C$) for a Gaussian line when added to the baseline continuum model. The blue, green, red contours represent values of $\Delta C$ equivalent to a confidence level of 99\%, 95\% and 68\%, respectively. 
Three emission lines are identified at a confidence level $>95\%$. 
\label{fig:bl_search}}
\end{figure}

Now we perform a blind search of emission/absorption lines in the co-added HEG and MEG spectra, following the method described in \citet{2010A&A...521A..57T} and \citet{2021NatAs...5..928S}. 
Specifically, the putative line is modelled by a Gaussian profile, whose centroid energy $E_0$ and normalization $N$ are free to vary, and the dispersion is fixed at half of the HETG resolution (0.0056 $\rm \AA$ for HEG and 0.0111 $\rm \AA$ for MEG).
The Gaussian is added to the baseline power-law model derived in Section \ref{subsec:cont} (i.e., with a fixed photon-index of 1.723), progressively increasing $E_0$ channel by channel over the energy range of interest and seeking for the best-fit $N$. 
The improvement of \emph{C}-stat, $\Delta C$, with respect to the baseline model is evaluated at each step.  
For each pair of ($E_0$, $N$), confidence level of the line can be calculated according to ${\Delta}C$, given the fact that $C$-stat is asymptotically distributed as $\chi^{2}$.
The resultant contour plots are shown in Figure \ref{fig:bl_search} for the HEG spectrum between 5--8 keV. 
%and Figure \ref{fig:blsearch_soft} for the MEG/HEG spectra between 1--5 keV.

Three emission lines, all with a confidence level $>95\%$, are identified from the HEG spectrum between 5--8 keV. We note that these lines were also identified by \citet{2008MNRAS.389L..52B} based on the same dataset.
The centroid energy of the three lines, after correcting for the systemic redshift ($z$=0.00584) of NGC\,7213, is 6.391 keV, 6.709 keV and 7.001 keV, respectively. 
We have checked the robustness of line detection by varying the photon-index of the baseline power-law model between 1.70--1.74. 
The resultant confidence levels of the three lines all remain above 95\%.
To refine the characterization of each line, we again apply the Gaussian model, this time setting the line width as a free parameter. The Gaussian fit results, including line centroid, line width, flux and equivalent width, are listed in Table \ref{tab:blsearch}. 
%shown in Figure \ref{fig:gauss_fit}a.

\begin{deluxetable}{CLCLLL}
%\tablenum{2}
\tablecaption{Spectral fit results of the Fe lines\label{tab:blsearch}}
\tablewidth{0pt}
\tablehead{
\colhead{$E_{\rm 0}$} & \colhead{Significance} & \colhead{$E_{c\rm }$} & \colhead{$\sigma_{\rm E}$} & \colhead{$F$} & \colhead{EW} 
%\colhead{$\rm (keV)$} & \colhead{ } & \colhead{$\rm (keV)$} & \colhead{$\rm (eV)$} & \colhead{$\rm (erg~s^{-1}~cm^{-2})$} & \colhead{$\rm (eV)$} 
}
\colnumbers
\startdata
%6.354 before z-corrected
6.391 & $>99.99\%$ & $6.395^{+0.007}_{-0.008}$ & $20^{+10}_{-8}$ & $3.1^{+0.8}_{-0.7}$ & $126.9^{+0.5}_{-0.8}$ \\
%6.670
6.709 & $96.7\%$ & $6.717^{+0.016}_{-0.013}$ & $<55$ & $0.7^{+0.5}_{-0.5}$ & $26.0^{+0.4}_{-0.2}$ \\
%6.691
7.001 & $>99.99\%$ & $6.996^{+0.012}_{-0.011}$ & $<99$ & $1.4^{+0.7}_{-0.6}$ & $61.2^{+0.8}_{-0.3}$ \\ % without Fe Kb line 
%7.001 & \gt99.99\% & 6.990^{+0.017}_{-0.002} & \lt65 & 1.5^{+0.7}_{-0.6}\times10^{-13} & 60.6^{+0.6}_{-0.4}\\ % with Fe Kb line
\enddata
% already correlated for the redshift of NGC7213
\tablecomments{(1) Most probable line centroid energy, in units of keV, determined by the blind search and corrected for the systemic redshift of NGC\,7213; (2) Significance level of the line; (3) Line centroid in units of keV, derived from the Gaussian fit; (4) Line width in units of eV; (5) Line flux in units of $10^{-13}\rm~erg~s^{-1}~cm^{-2}$; (6) Equivalent width in units of eV. Error bars are given at 90\% confidence level and upper limits of the line width are given at 3\,$\sigma$. }
\end{deluxetable}

\begin{figure*}[htb!]
%\gridline{\fig{fit_4gauss_bin1_20211207.eps}{0.49\textwidth}{}
%\fig{fit_apec_20211207.eps}{0.49\textwidth}{}
%}
\centering\includegraphics[width=0.49\textwidth]{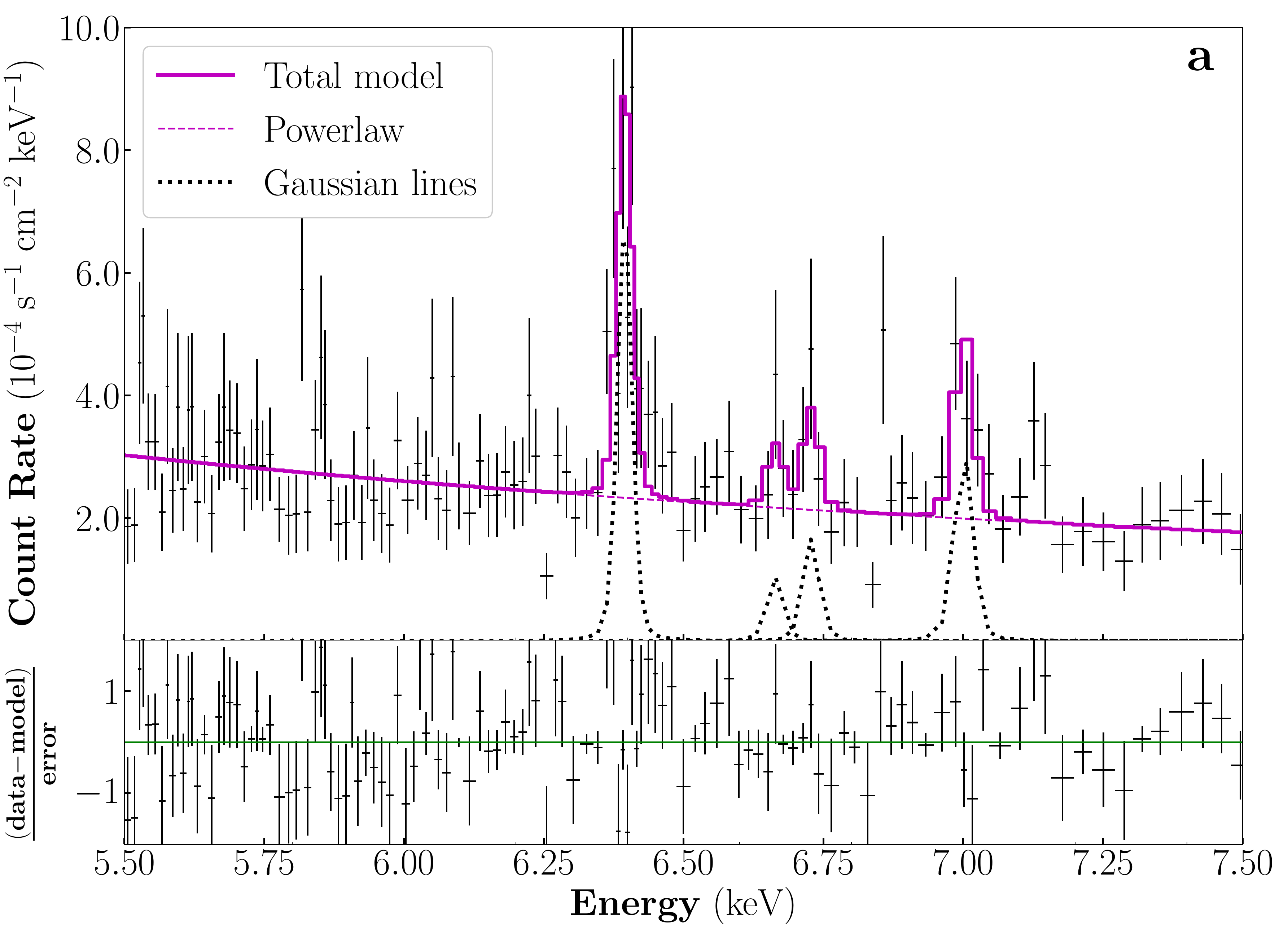}
\centering\includegraphics[width=0.49\textwidth]{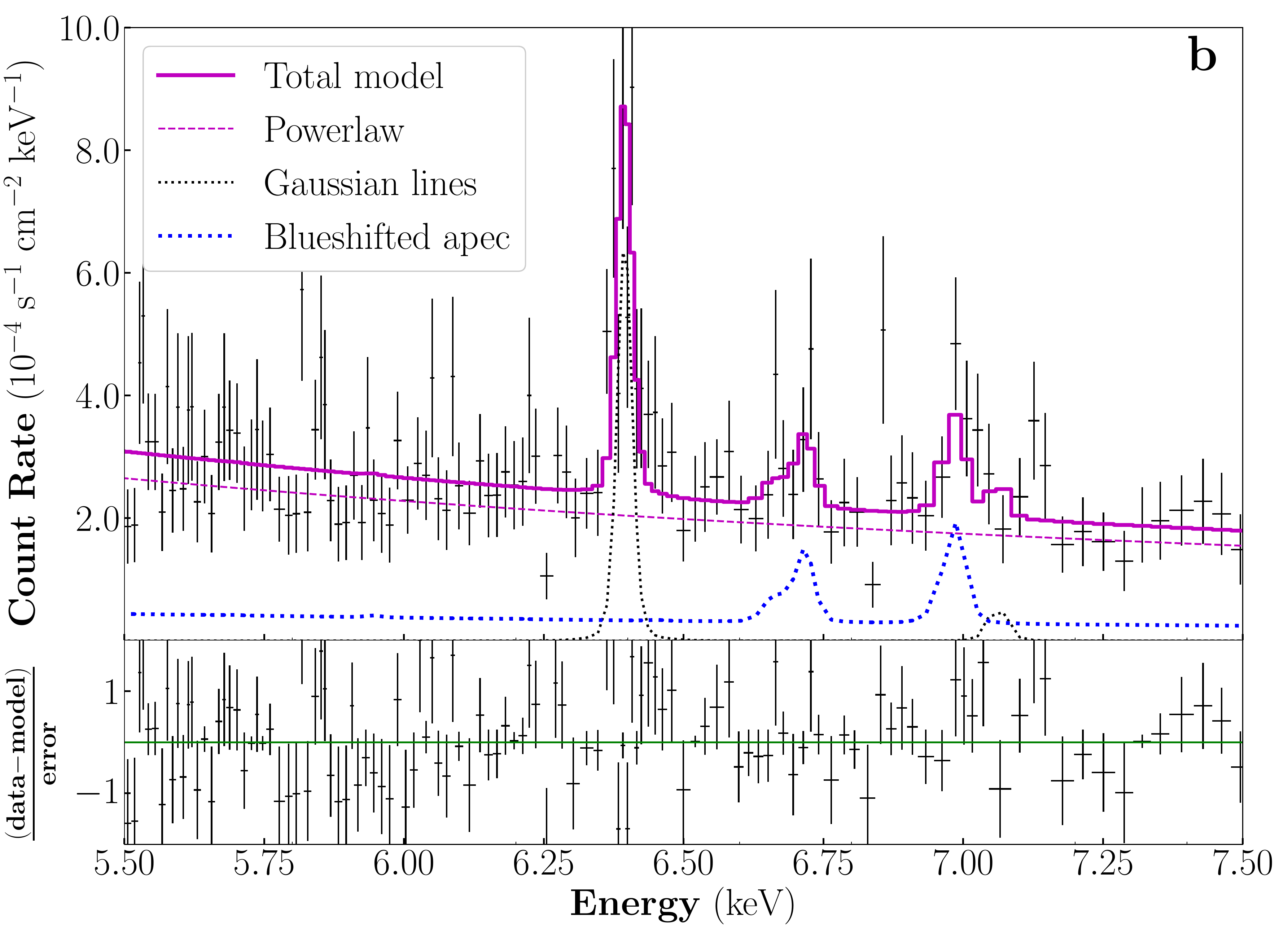}
\caption{(a) The co-added HEG spectrum (black data points) fitted with a baseline power-law (magenta dotted line) plus four Gaussian lines (black dotted curves) representing, from low to high energies, Fe I K$\alpha$, Fe XXV K$\alpha$ (forbidden transition), Fe XXV K$\alpha$ (resonant transition) and Fe XXVI Ly$\alpha$. The total model is represented by the magenta solid curve.  The spectrum is binned to have a signal-to-noise ratio greater than 3 for better illustration. The error bars are of 1\,$\sigma$. The lower panel gives the residual-to-error ratio. (b) Similar to panel (a), but the fitted model consists of the baseline power-law, a blueshifted \emph{apec} (blue dotted curve) and two Gaussian lines representing Fe I K$\alpha$ and K$\beta$.
\label{fig:gauss_fit}}
\end{figure*}

The 6.391 keV line can be identified as the K$\alpha$ line of neutral or weakly ionized Fe.
%centered on $6.395^{+0.007}_{-0.008}\rm~keV$ should consist of a doublet but cannot be well resolved by HETG. 
This line is marginally resolved, with a best-fit line width of $20^{+10}_{-8}$ eV, corresponding to a velocity dispersion of $1.0^{+0.4}_{-0.4}\times10^3\rm~km~s^{-1}$.
The relatively low velocity dispersion and lack of significant Doppler shift suggest that this line arises from a truncated accretion disk or remote cold gas illuminated by the central LLAGN. A similar conclusion was drawn by \citet{2010MNRAS.408..551L}. 
%Neutral iron fluorescence line could either be generated in illuminated distant cold material or reflected by the rotating thin accretion disk outside the truncation radius. For the latter case, the velocity dispersion of iron K$\alpha$ line indicates its origin position is $\sim10^5\rm~r_g$ away from the central SMBH.   
The 6.709 keV line is the weakest among the three lines and is consistent with the Fe XXV (He-like Fe) K$\alpha$ triplet, which is not fully resolved by HEG. 
The 7.001 keV line may be %$7.002^{+0.001}_{-0.019}\rm~keV$ has been identified as 
a blueshifted Fe XXVI (H-like Fe) Ly$\alpha$ line, as previously suggested by \citet{2008MNRAS.389L..52B}.
For the rest-frame energy of Fe XXVI Ly$\alpha$\footnote{Rest-frame line centroid energies throughout this work are taken from ATOMDB (http://www.atomdb.org).}, $6.966\rm~keV$, this implies for a line-of-sight velocity of $\sim -1.3\times10^3\rm~km~s^{-1}$. 
We further consider the possibility that this line is Fe I K$\beta$, which has a rest-frame energy of 7.058 keV. However, this would imply for a line-of-sight velocity of $\sim2400\rm~km~s^{-1}$, which is inconsistent with the observed centroid of Fe I K$\alpha$. Moreover, the flux ratio between Fe I K$\beta$ and K$\alpha$ should have a canonical value of $\sim0.13$ \citep{2003A&A...410..359P}, while the observed ratio between the 7.001 keV and 6.391 keV lines is substantially larger ($0.4\pm0.2$). Hence we conclude that the 7.001 keV line cannot be associated with Fe I K$\beta$. 
Guided by the implied blueshift of the Fe XXVI Ly$\alpha$ line, we refit the blended Fe XXV K$\alpha$ using a double Gaussian profile (Figure~\ref{fig:gauss_fit}a), obtaining line centroids at $6.660_{-0.019}^{+0.015}\rm~keV$ and $6.722_{-0.016}^{+0.017}\rm~keV$. 
If the two Gaussians could be respectively associated with the forbidden transition (rest-frame energy at 6.637 keV) and resonant transition (rest-frame energy at 6.700 keV), they would both be blueshifted with a line-of-sight velocity of $\sim -1.0\times10^3\rm~km~s^{-1}$, nicely compatible with the putative blueshifted Fe XXVI Ly$\alpha$. This lends further support to the identification of the latter. 
%We further applied a gaussian line centered on the putative energy of intercombination Fe XXV K$\alpha$ line with the same velocity shift.
%And the flux ratio between Fe XXV \emph{f}-line and \emph{r}-line has been constrained to $0.5^{+0.9}_{-0.4}$ through bootstrapping, which is consistent with the 
%And the G-ratio $G=\frac{f+i}{r}$ of the He-like iron triplet has been constrained to an 2-$\sigma$ upper limit of $G\lesssim5.3$.
%\textbf{(i-line only has  upper limit)}. 

\begin{deluxetable}{lcccccc}[hb!]
%\footnotesize
%\tablenum{3}
\tablecaption{Additional significant lines\label{tab:softband}}
\tablewidth{0pt}
\tablehead{
\colhead{Line identification} & \colhead{$E_{\rm lab}$} & \colhead{Significance} & \colhead{$E_{\rm c}$} & \colhead{$\sigma_{\rm E}$} & \colhead{$F$} & \colhead{EW}\\
\colhead{(1)} & \colhead{(2)} & 
\colhead{(3)} & \colhead{(4)} & \colhead{(5)} & \colhead{(6)} & \colhead{(7)}
}
%\colnumbers
\startdata
%\multirow{2}*{Ne X Ly$\alpha$} & \multirow{2}*{1.022} & 94.5\% & 1.019^{+0.004}_{-0.004} & - & 9^{+6}_{-6} & 1.2^{+0.2}_{-0.2} \\
%{             } & {   } & - & - & - & \textless17 & \textless1.8 \\
%\hline
%\multirow{2}*{Na K$\alpha$} & \multirow{2}*{1.041} & - & 1.038^{+0.010}_{-0.020} & - & \textless13 & \textless2.9 \\
%{          } & {   } & 95.8\% & 1.040^{+0.006}_{-0.003} & \textless9 & 12^{+9}_{-7} & 1.4^{+0.1}_{-0.1} \\
%\hline
%\multirow{2}*{Ne X Ly$\beta$} & \multirow{2}*{1.211} & 93.0\% & 1.207^{+0.006}_{-0.011} & \textless8 & 5^{+4}_{-4} & 0.64^{+0.05}_{-0.04} \\
%{           } & {   } & - & - & - & \textless16 & \textless1.1 \\
%\hline
\multirow{2}*{Mg XII Ly$\alpha$} & \multirow{2}*{1.473} & 97.2\% & $1.471^{+0.006}_{-0.006}$ & $<71$ & $0.7^{+0.4}_{-0.4}$ & $1.1^{+0.1}_{-0.1}$ \\
%\multirow{2}*{Mg XII Ly$\alpha$} & \multirow{2}*{1.473} & 97.2\% & \multirow{2}*{1.471} & <71 & 7^{+4}_{-4} & 1.1^{+0.1}_{-0.1} \\
{               } & {   } & - & 1.471 & - & $<1.2$ & $0.97^{+0.01}_{-0.01}$ \\
%{               } & {   } & - & - & - & \textless12 & 0.97^{+0.01}_{-0.01} \\
\hline
%\multirow{2}*{Si XIII K$\alpha$ (f)} & \multirow{2}*{1.839} & - & 1.840^{+0.013}_{-0.013} & - & \textless13 & 0.67^{+0.01}_{-0.03} \\
%{                   } & {   } & - & 1.84^{+0.19}_{-0.19} & - & <11 & 0.19^{+0.01}_{-0.01} \\
%\hline
%\multirow{2}*{Si XIII K$\alpha$ (i)} & \multirow{2}*{1.854} & - & - & - & \textless10 & \textless0.13 \\
%{                   } & {   } & 94.5\% & 1.856^{+0.009}_{-0.009} & \textless13 & 7^{+7}_{-6} & 1.12^{+0.03}_{-0.01}\\
%\hline
%\multirow{2}*{Si XIII K$\alpha$ (r)} & \multirow{2}*{1.865} & - & - & - & \textless13 & 0.45^{+0.01}_{-0.03} \\
%{                   } & {   } & - & 1.867^{+0.009}_{-0.014} & - & \textless20 & 0.93^{+0.01}_{-0.03} \\
%\hline
\multirow{2}*{Ar XVIII Ly$\alpha$} & \multirow{2}*{3.321} & 96.9\% & $3.325^{+0.019}_{-0.010}$ & $<40$ & $2.3^{+1.8}_{-1.6}$ & $5.83^{+0.04}_{-0.10}$\\
%{                 } & {   } & - & 3.32^{+0.03}_{-0.08} & - & <55 & 2.6^{+0.1}_{-0.1} \\
%\multirow{2}*{Ar XVIII Ly$\alpha$} & \multirow{2}*{3.321} & 96.9\% & \multirow{2}*{3.326} & <40 & 23^{+18}_{-16} & 5.83^{+0.04}_{-0.10}\\
{                 } & {   } & - & 3.325 & - & $<5.5$ & $2.6^{+0.1}_{-0.1}$ \\
\hline
\multirow{2}*{Ca XIX K$\alpha$ (r)} & \multirow{2}*{3.902} & - & 3.911 & - & $<5.5$ & $4.4^{+0.2}_{-0.2}$ \\
{              } & {   } & 96.1\% & $3.911^{+0.004}_{-0.003}$ & $<11$ & $2.5^{+1.8}_{-1.6}$ & $7.3^{+0.2}_{-0.1}$\\
\enddata
% already correlated for the redshift of NGC7213
\tablecomments{(1)-(2) Identified line transition and rest-frame centroid energy in units of keV. Each line has two rows, the upper row for the MEG spectrum and the lower row for the HEG; (3) Significance of the line in the blind search; (4) Line centroid in units of keV, derived from the Gaussian fit; (5) Line width in units of eV; (6) Line flux in units of $10^{-14}\rm~erg~s^{-1}~cm^{-2}$; (7) Equivalent width in units of eV. Errors are quoted at 90\% confidence level, while all the upper limits are of 3\,$\sigma$. 
%Instrumental broadening alone is considered for narrow lines that line width fitting cannot converge or peg at lower limits, denoted with '-'. 
}
\end{deluxetable}
\begin{figure}[ht!]
\centering\includegraphics[width=1\textwidth]{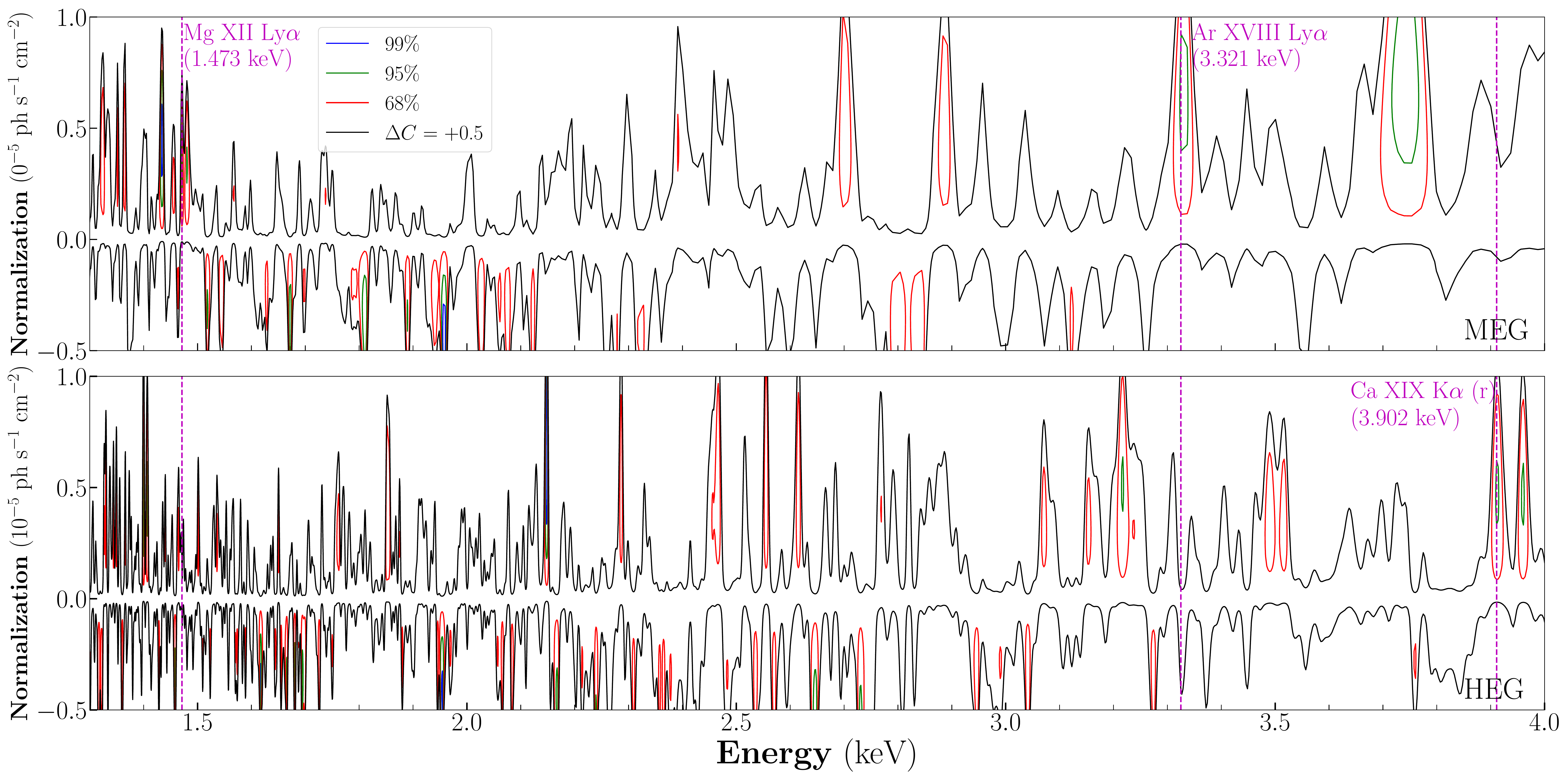}
%\centering\includegraphics[width=0.8\textwidth]{blind_search_2-4_20211022.pdf}
\caption{Blind search of lines in the co-added MEG (upper panel) and HEG (lower panel) spectra between 1--4 keV. Contours denote the improvement of $C$-stat (${\Delta}C$) for a Gaussian line when
added to the baseline continuum model. The blue, green, red contours represent values of ${\Delta}C$ equivalent to a confidence level of 99\%, 95\% and 68\%, respectively. An emission line is considered real if its significance were greater than 95\% in at least one of the two spectra and their flux or 3\,$\sigma$ upper limit were consistent between the two spectra. Three significant lines are thus detected and labeled with the tentatively identified atomic transition. 
\label{fig:blsearch_soft}}
\end{figure}

%{\bf (bootstrap test 4 lines significance)[to be done]}
Our blind search also finds several significant emission lines between 1--4 keV in the MEG and HEG spectra (Figure~\ref{fig:blsearch_soft}), where the two gratings have a comparable line sensitivity. 
For these lines to be considered {\it real}, we have required that they are detected at a significance above 95\% in at least one of the two spectra and that their flux or 3\,$\sigma$ upper limit be consistent between the two spectra. 
The basic properties of these lines are again derived from a Gaussian fit and listed in Table \ref{tab:softband}. 
From the line centroid, we identify Ly$\alpha$ lines of Mg XII and Ar XVIII. Both lines are unresolved and have a Doppler shift  $\lesssim450\rm~km~s^{-1}$. 
%In addition, a line at 1.207 keV is identified with the Ly$\beta$ of Ne X.
We also tentatively associate an unresolved 3.911 keV line with the resonant transition of Ca XIX K$\alpha$ (rest-frame energy 3.902 keV), which would imply for a Doppler shift $\lesssim700\rm~km~s^{-1}$. 
%Finally, a line at 1.856 keV line is tentatively identified as the intercombination line of Si XIII K$\alpha$, with the resonant and forbidden lines among the triplet being marginally detected.
%These H-like and He-like lines may be associated with a barely static hot plasma after correcting for the systemic redshift (z=0.00584).
%There is no significant fluorescence K$\alpha$ line from neutral atoms except for a likely Na K$\alpha$ line at 1.040 keV.
We note that no significant emission lines are found between 4--6 keV. Nor do we find any significant absorption line in the MEG and HEG spectra.

\subsection{Modelling the hot plasma} \label{subsec:spec}
The blueshifted Fe XXVI Ly$\alpha$ and Fe XXV K$\alpha$ lines indicate a hot plasma with a substantial bulk motion. We use a phenomenological model to characterize such a plasma. Specifically, an optically-thin plasma model, {\it apec} in Xspec, is applied to fit the 5--8 keV HEG spectrum, with the plasma temperature and redshift as free parameters and the abundance fixed at solar. 
The portion of the spectrum at lower energies is temporarily neglected, which may require an additional component as suggested by the emission lines of the low-$Z$ elements and a small excess to the power-law continuum seen at energies $\lesssim$1 keV in the MEG spectrum (Figure~\ref{fig:baseline}).
Foreground absorption is again fixed at the Galactic value.  
Two Gaussian lines are added to account for the Fe I K$\beta$ and K$\alpha$ lines, with a fixed line ratio of 0.13 between the two.
The baseline power-law continuum is also included, with the photon-index fixed at 1.723.
This composite model results in a reasonable fit (Figure~\ref{fig:gauss_fit}b), with a plasma temperature $kT = 16\pm6\rm~keV$, which is mainly driven by the flux ratio between the Fe XXVI Ly$\alpha$ and Fe XXV K$\alpha$ lines. 
%Best-fit results can be referred to Table \ref{tab:mod_fit} and Figure \ref{fig:gauss_fit}(b). 
After correcting for the systemic redshift of NGC\,7213, the fitted model requires a line-of-sight velocity of $-1.1^{+0.3}_{-0.4}\times10^3\rm~km~s^{-1}$, consistent with the expectation from the line centroids (Section~\ref{subsec:blsearch}). 
The unabsorbed 2--10 keV of this 16-keV plasma is found to be $\sim4.0\times10^{41}\rm~erg~s^{-1}$. 

While the HEG spectrum shows no sign of a redshifted counterpart to the blueshifted Fe lines, it is tempting to provide constraint on a potential redshifted component. 
To do so, we add to the above composite model a second \emph{apec} component, requiring it to have the same temperature and a line-of-sight velocity exactly opposite to the blueshifted component. 
The upper limit of this redshifted \emph{apec} is then determined by deviating the $C$-stat until the 3$\sigma$ level is arrived at. The 2--10 keV flux ratio between the redshifted and blueshifted components is thus constrained to be  $<$0.45.
%$1.8\times10^{41}\rm~erg~s^{-1}$
%Detailed explanation for the lack of red-shifted component will be discussed in Section \ref{subsec:scenario}.

The 16-keV plasma, however, cannot be responsible for the detected K$\alpha$ and Ly$\alpha$ emission lines of the low-$Z$ elements (Table~\ref{tab:softband}), since at such a high temperature these low-$Z$ elements would be fully ionized. 
%Adopting the same models to simultaneously fit HEG (1--8 keV) and MEG (0.5--5 keV) spectra reveals a soft excess between 0.5-1 keV band. 
To account for these lines,  
we again add a second {\it apec} component to the above composite model, this time allowing both the plasma temperature and redshift to vary freely.
It turns out that this new composite model provides a reasonable simultaneous fit to the 1--8 keV HEG spectrum and 0.5--5 keV MEG spectrum. 
Compared to the single power-law model, the composite model improves the $C$-stat by 80 for the HEG spectrum and 10 for the MEG spectrum for 6 less degree-of-freedom.
The best-fit temperature of the second {\it apec} is $0.8_{-0.1}^{+0.2}\rm~keV$,
and the best-fit redshift is nearly zero, suggesting little bulk motion.
The unabsorbed 0.5-10 keV luminosity of this component is $8.7\times10^{39}\rm~erg~s^{-1}$.
This signifies a low-$T$ plasma that might be associated with a circumnuclear hot gas.
It is noteworthy that with the addition of the low-$T$ component, the best-fit temperature of the blueshifted high-$T$ component becomes $13^{+5}_{-2}$ keV, still consistent with the value (16 keV) determined from fitting the 5--8 keV spectrum alone.

The spectral fit results in this section are summarized in Table \ref{tab:mod_fit}.

\subsection{Zeroth-order spectrum of the off-nucleus region} \label{subsec:circum_reg}
\begin{figure}[ht!]
\centering\includegraphics[width=0.6\textwidth]{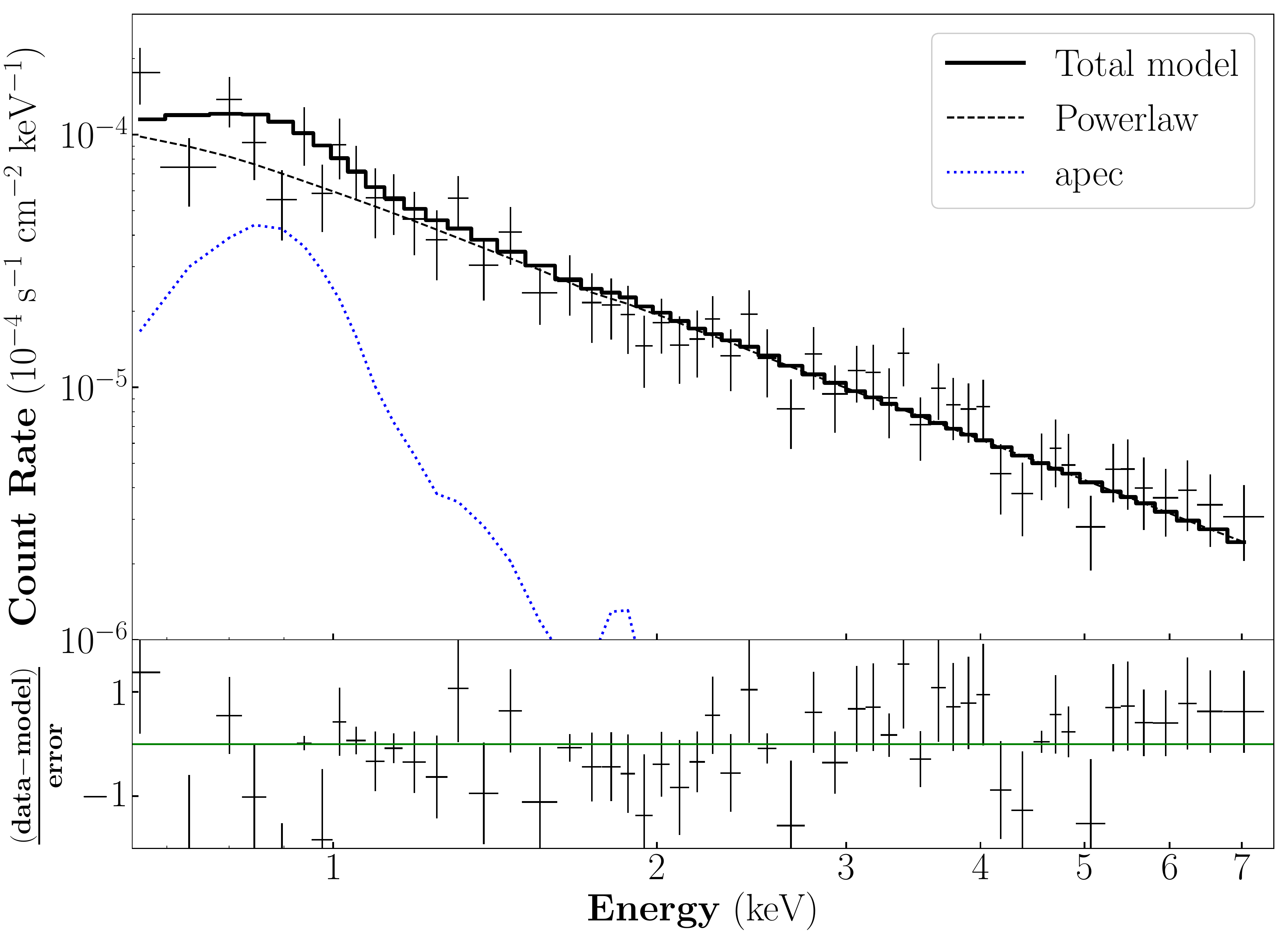}
\caption{Co-added zeroth-order spectrum of the off-nuclear region in NGC\,7213. 
The spectrum is binned to have a signal-to-noise ratio greater than 3 for better illustration.
The spectrum is fitted by an absorbed power-law (black dashed line) plus a maximally-allowed \emph{apec} model (blue dotted curve). The black solid curve marks the total model. The lower panel shows the residual-to-error ratio. Error bars are of 1\,$\sigma$. 
\label{fig:offcenter}}
\end{figure}

We have also examined the X-ray spectrum of an off-nucleus region (Figure~\ref{fig:offcenter}), which is extracted from the zeroth-order ACIS/HETG image and provides meaningful constraint on the spatial extent of the thermal components found in the first-order dispersed spectra.
The off-nucleus region is defined as an annulus with an inner-to-outer radius of 2.5\arcsec--5\arcsec\ (275--550 pc; see the insert in Figure~\ref{fig:merge}), which is immediately outside the first-order spectral extraction region 
and also avoids potential effect of pileup caused by the bright nucleus.
The corresponding background spectrum is extracted from a concentric annulus with an inner-to-outer radius of 5\arcsec--8\arcsec. 
%Several discrete sources present in the source and background regions are excluded.
Spectra extracted from the two {\it Chandra} observations are co-added.
We find that a power-law model subjected to the Galactic foreground absorption can well describe this spectrum. 
The best-fit photon index of $1.7\pm0.1$ and an unabsorbed 0.5--10 keV luminosity of $2.5\times10^{40}\rm~erg~s^{-1}$
can be understood as the collective emission from unresolved X-ray binaries plus PSF-scattered photons from the bright nucleus.

While a thermal component is not formally required by the apparently featureless spectrum, we add an \emph{apec} model to the fit, fixing the abundance at solar and the plasma temperature at $0.8$ keV as found for the low-$T$ component in the first-order spectra (Section~\ref{subsec:spec}).
The maximum normalization of this \emph{apec} is found such that the deviation of $C$-stat arrives at the 3\,$\sigma$ level, which is indicated by the blue dotted curve in Figure \ref{fig:offcenter}.
The corresponding unabsorbed 0.5--10 keV luminosity is $\sim4\times10^{38}\rm~erg~s^{-1}$, which is $\sim$20 times lower than that of the 0.8 keV component detected in the first-order spectrum. 
In terms of surface brightness, the 0.8 keV plasma is a factor of 70 times lower in this off-nucleus region, considering the underlying projected area of zeroth- and first-order spectra. This indicates that the 0.8 keV plasma is concentrated within the central $\sim$300 pc and strengthens our above suggestion that it is tracing a circumnuclear hot gas.

\section{A hot wind from the LLAGN as the origin of the Fe lines} 
\label{sec:wind}
\begin{figure}[ht!]
\centering\includegraphics[width=0.4\textwidth]{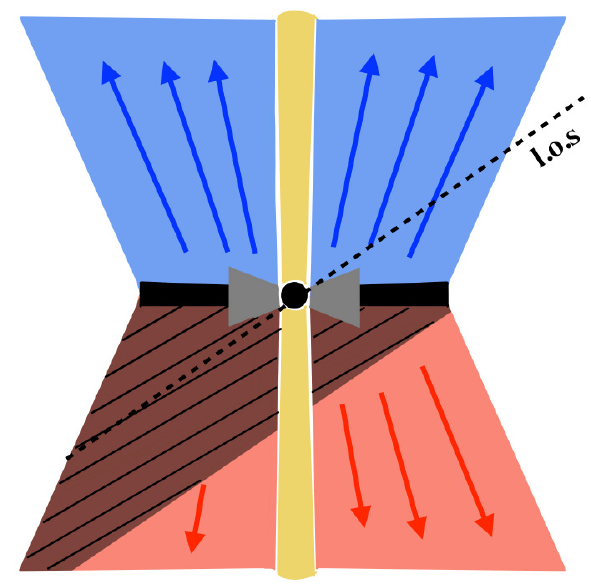}
\caption{Schematic diagram of the inflow-outflow system around the central SMBH in NGC\,7213 (represented by the black circle). The pair of grey triangles represent the geometrically-thick hot accretion flow. The black stripes along the equatorial plane mark the truncated thin disk, while the vertical yellow stripe denotes a relativistic jet. The blue and red cones with large opening angles represent the hot wind propagating towards and away from the observer, where emission lines of highly-ionized Fe are generated. In the case of NGC 7213, a high viewing angle (defined as away from the jet axis) is inferred, such that a large fraction of the redshifted hot wind is obscured by the truncated disk.
%For M81, line-emission region remains visible with a small viewing angle and larger truncation radius.
\label{fig:AD_absorb}}
\end{figure}

%\subsection{Physical scenario} %\label{subsec:scenario}
The X-ray spectra presented in the previous section reveal the presence of a high-velocity, hot plasma in the LLAGN of NGC\,7213. 
This is highly similar to the case of M81$^*$, in which a high-velocity, hot plasma is traced by highly-ionized Fe lines and interpreted as an energetic wind driven by the hot accretion flow onto the central SMBH \citep{2021NatAs...5..928S}.
Here we consider this same scenario for NGC\,7213, which is illustrated in Figure \ref{fig:AD_absorb}.
In this schematic diagram of an inflow-outflow system around a weakly accreting SMBH, a cold thin disk dictates the accretion inflow at large radii, but becomes truncated at a radius $r_{\rm tr} \sim 10^3~r_{\rm g}$. Inside $r_{\rm tr}$, the disk is replaced by a geometrically-thick hot inflow. 
Close to the event horizon, a highly-collimated, relativistic jet is expected to launch.  
%Both theory \cite{1999MNRAS.303L...1B} and numerical simulations of hot accretion flows \citep{2012ApJ...761..130Y,2012ApJ...761..130Y,2015ApJ...804..101Y} predict that 
Also symbiotic with the hot accretion flow is a bipolar, non-relativistic wind, which ultimately carries away the bulk of the inflowing mass.
Compared to the jet, the wind originates from a wide range of radius in the accretion flow, having a much wider opening angle and much smaller radial velocity.
Due to adiabatic expansion, the wind has its temperature decreasing from that of the hot accretion flow and spanning a range well suited for producing the highly-ionized Fe lines.
We are thus motivated to perform magnetohydrodynamical (MHD) simulations of the putative wind, tailored to the physical conditions of the SMBH in NGC\,7213, to test this favorable origin of the observed Fe lines.
%Either the truncated thin disk or the hot accretion flow could generate fast-moving outflow. Only the latter one can drive wind as hot as $\sim10$ keV and produce emission in X-ray regime.
%Similar to the case of M81$^*$, we considered the fast-moving hot plasma could be evidence of energetic hot wind generated from hot accretion inflow. 
%As shown in the schematic diagram in Figure \ref{fig:AD_absorb}, blue-shifted highly ionized iron lines could be generated in the branch of hot wind coming toward us.

\subsection{Wind simulation and synthetic wind spectrum}
\label{subsec:simulation}
We perform a 2.5-dimension MHD simulation with the ZEUS-MP/2 code version 1.5.19 \citep{2006ApJS..165..188H} in spherical coordinates. 
The black hole mass is set to be $10^8\rm~M_{\odot}$ based on observations \citep{2002ApJ...579..530W}. 
The simulation setup is similar to  \citet{2021NatAs...5..928S}, which is briefly described here.
The wind starts from the inner boundary, which is set at 2000 $r_{\rm g}$, i.e., the assumed disk truncated radius, appropriate for the moderate Eddington ratio of the LLAGN in NGC\,7213 and also consistent with the narrowness of the Fe I K$\alpha$ line (Section~\ref{subsec:blsearch}).
We ignore a possible wind from the cold disk outside the truncated radius, which is expected to have too low temperatures to produce the highly ionized Fe.
The rotational energy, kinetic energy, thermal energy and magnetic energy of the wind are taken into account. 
Following \citet{2015ApJ...804..101Y} and \citet{2021ApJ...914..131Y}, we set the rotation velocity as $0.8v_{\rm k}$, wind velocity as $0.6v_{\rm k}$, temperature as $0.5kT_{\rm vir}$, and the plasma $\beta=p_{\rm gas}/p_{\rm mag}=0.2$ at the inner boundary, where $v_{\rm k}=(GM_{\rm BH}/r)^{0.5}$, $kT_{\rm \rm vir}=GM_{\rm BH}/(3r)$, $p_{\rm gas}$, and $p_{\rm mag}$ are the Keplerian velocity, virial temperature, gas pressure and magnetic pressure, respectively.
Only the toroidal magnetic field is taken into account, since the poloidal field is typically much smaller than the toroidal component at large radii as found in wind simulations \citep{2021ApJ...914..131Y}. The grid of the simulation is extended from $2\times10^3~r_{\rm g}$ to $10^6~r_{\rm g}$ in radial direction. A radially injecting boundary condition is set for the inner boundary, while an outflow boundary condition is adopted at the outer boundary. 
The grid in latitudinal direction is extended from $\theta = 10^{\circ}$ to $50^{\circ}$ to avoid the singularity of spherical coordinate and to neglect the jet. The existence of a relativistic jet does not affect the dynamics of the wind, nor does it produce emission lines.
A reflective boundary condition is set for the latitudinal direction, since we expect that the pressure from the truncated disk will confine the hot wind.
\begin{figure*}[ht!]
%\gridline{\fig{temp_distribution_20211124.eps}{0.49\textwidth}{}
\centering\includegraphics[width=0.49\textwidth]{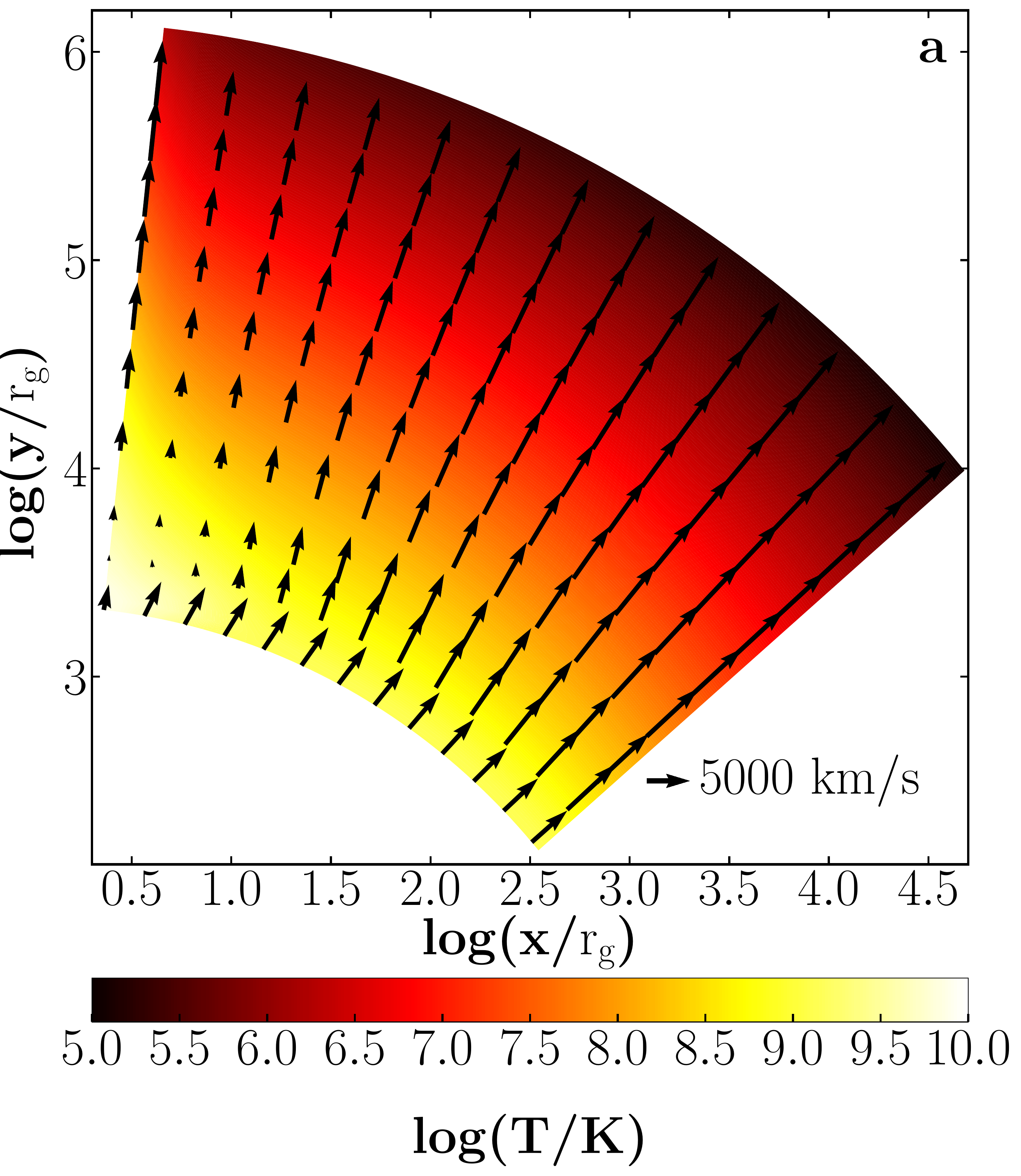}
\centering\includegraphics[width=0.485\textwidth]{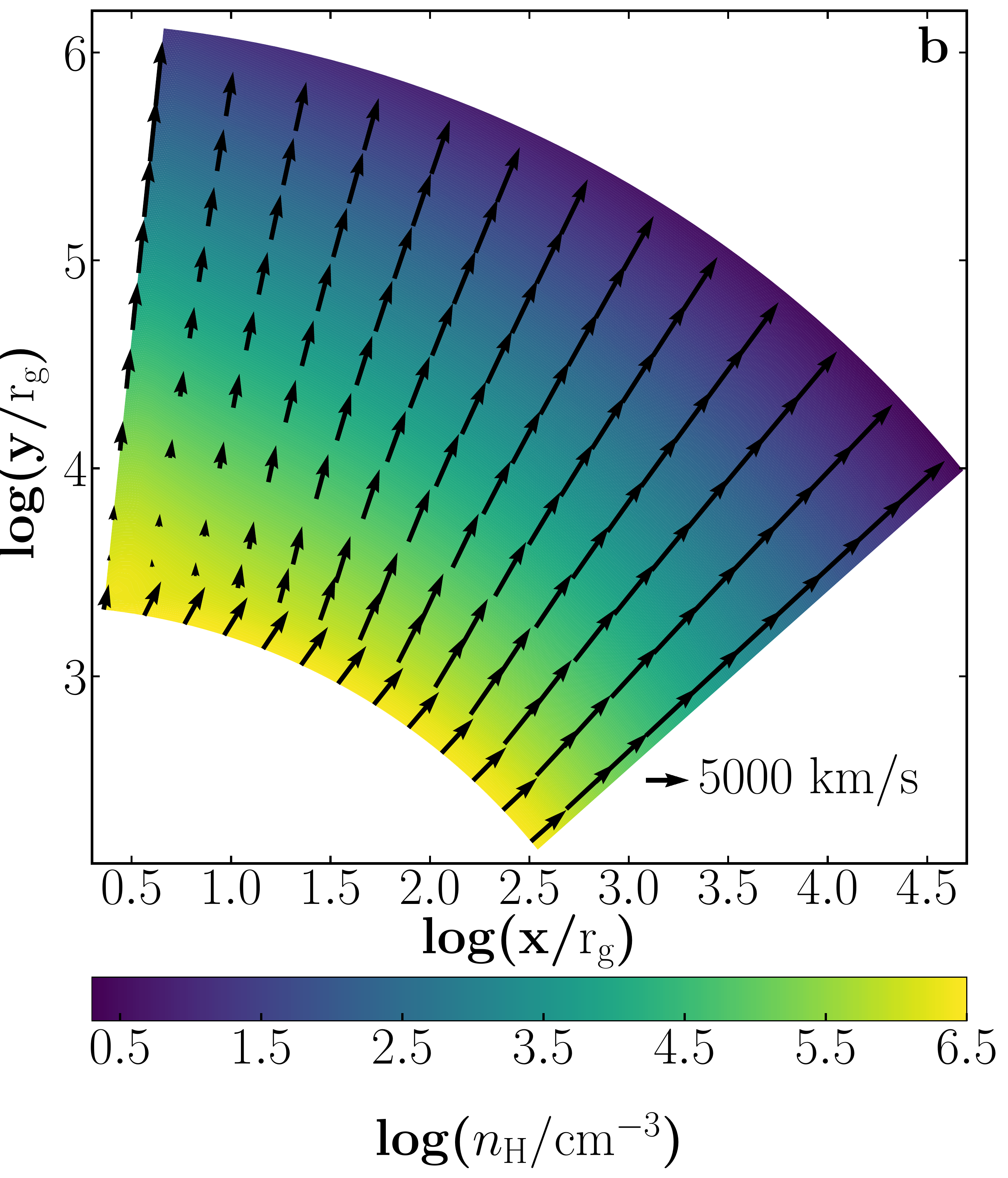}
%\fig{density_distribution_20211124.eps}{0.485\textwidth}{}
%}
\caption{Plasma temperature (a) and density (b) distribution of the hot wind from the MHD simulation. The radial velocity field is overlaid with arrows. The y-axis follows the jet axis and the x-axis defines the equatorial plane of the truncated disk. Regions occupied by the jet or the disk have been excluded.
\label{fig:distri}}
\end{figure*}

A direct outcome of the simulation is the two-dimensional (2D) distributions of density, temperature and velocity, as shown in Figure~\ref{fig:distri}. 
We note that the absolute scale of the density is determined by matching the synthetic wind spectrum to the observed spectrum (see Section~\ref{subsec:synthetic}). 
The synthetic wind spectrum is produced based on the simulated  quantities. 
Specifically, we first expand the 2D grid into a 3D grid, given the axisymmetry about the jet axis and mirror symmetry with respect to the equatorial plane.
We then calculate the density-weighted X-ray spectrum of each grid in the black hole rest frame, utilizing ATOMDB for a given metal abundance to include atomic transitions and bremsstrahlung. 
The synthetic spectrum is then calculated by integrating along a given viewing angle, i.e., the angle between the jet axis and line-of-sight. 
In this step we take into account the effect of gravitational lensing, which will boost the redshifted component, and the effect of relativistic Doppler shift, which will boost (reduce) the blueshifted (redshifted) component.
Self-absorption in the hot wind is neglected, since the hot wind is expected to be optically-thin, which we verify in Appendix~\ref{sec:depth}. However, we have considered potential absorption by the truncated thin disk, which could be optically-thick to X-ray photons from behind the disk, for certain light-of-sight (see illustration in Figure~\ref{fig:AD_absorb}).

\subsection{Comparison with the observed HEG spectrum}
\label{subsec:synthetic}
%By matching the synthetic hot wind spectrum with the observed one, we could derive the normalization of the absolute gas density. The metal abundance of the wind was set to be twice as solar abundance.
The synthetic wind spectrum is compared with the observed HEG spectrum over 5--8 keV.
Added to the wind model are again the baseline power-law and two Gaussian lines representing Fe I K$\alpha$ and K$\beta$.
We emphasize that this comparison is not a formal fit to the observed spectrum. Rather, our aim is to examine whether the synthetic spectrum of the hot wind, naturally arising from the standard paradigm of hot accretion flow onto a weakly accreting SMBH, can provide a satisfactory description to the observed spectrum, in particular the blueshifted Fe emission lines. 

\begin{figure}[hb!]
%\centering\includegraphics[width=0.49\textwidth]{figures/fit_hotwind_a1_blue_211208.eps}
\centering\includegraphics[width=0.6\textwidth]{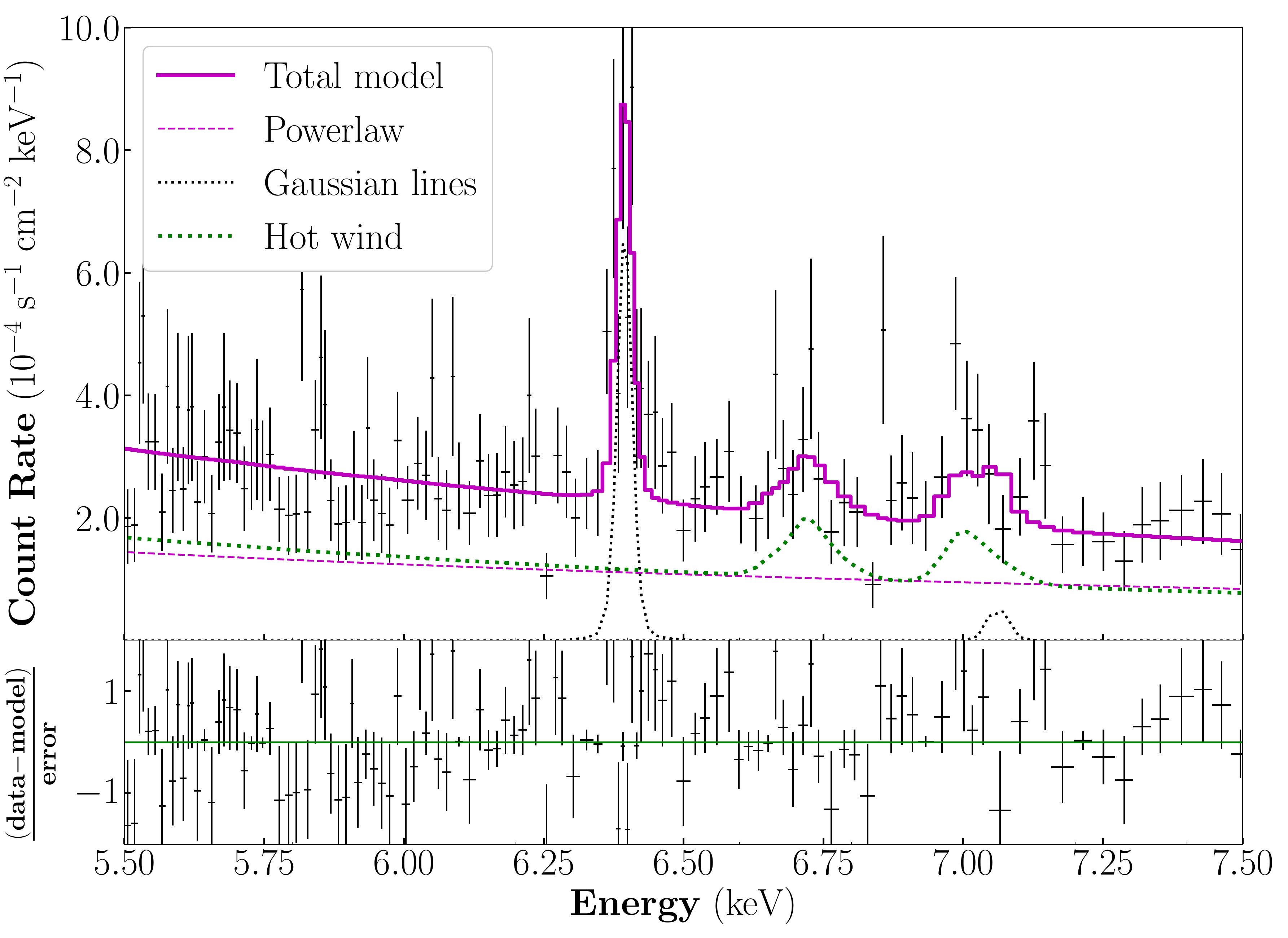}
\caption{The synthetic wind spectrum (green dotted curve) for a viewing angle of 80$^\circ$ and a Fe abundance of twice solar is confronted  with the observed spectrum. Other symbols are the same as in Figure~\ref{fig:gauss_fit}.
\label{fig:wind}}
\end{figure}

By construction of our wind simulation and synthetic spectrum, the primary free parameter is the viewing angle. We find that a viewing angle of 80$^{\circ}$ results in a synthetic spectrum that best matches the observed spectrum (Figure \ref{fig:wind}). 
In this case, the bulk of the highly-ionized Fe lines are produced in the radial range of $10^4-10^5~r_{\rm g}$, where the plasma temperature ranges between $3\times10^7-3\times10^8$ K and the radial velocity ranges between 1500--8000$\rm~km~s^{-1}$ (Figure~\ref{fig:distri}). 

The large viewing angle has two main effects. First, the projected line-of-sight velocity of the Fe line prevalent region takes values of $\sim1100\rm~km~s^{-1}$, compatible with the observed Doppler shift. 
Second, the redshifted component, which arises from behind the equatorial plane, is largely obscured by the truncated disk. 
This breaks the intrinsic symmetry between the blueshifted and redshifted components as a result of the intrinsic symmetry of the bipolar wind about the equatorial plane by design, and provides a natural explanation to an unseen redshifted counterpart to the blueshifted Fe XXVI Ly$\alpha$ line in the observed spectrum (Section~\ref{subsec:spec}).
Quantitatively, the thin disk has a large hydrogen column density of $4.3\times10^{26}\left(\frac{\alpha}{0.1}\right)^{-4/5}\left(\frac{M_{\rm BH}}{10^8\rm~M_{\odot}}\right)^{1/5}\left(\frac{\dot{m}}{0.001}\right)^{7/10}\left(\frac{r}{2000~r_{\rm g}}\right)^{-3/4}\rm~cm^{-2}$ along the direction perpendicular to the disk plane, for the black hole mass $M_{\rm BH}$, the accretion rate normalized by the Eddington limit $\dot{m}$, and viscosity parameter $\alpha$ suitable for NGC\,7213 \citep{2008bhad.book.....K}. 
The corresponding effective optical depth of electron scattering,  $\tau_{*}\sim4.7\times10^3(\frac{\alpha}{0.1})^{-4/5}(\frac{M_{\rm BH}}{10^8\rm~M_{\odot}})^{1/5}(\frac{\dot{m}}{0.001})^{1/5}$, is sufficiently high to block the X-ray photons from behind the disk. 

Alternatively, the weak or absent redshifted component may be due to an intrinsic asymmetry of the hot wind. 
%However, the bipolar nature of the energetic hot wind predicts the existence of a pair of blue-shifted and red-shifted hot plasma with similar velocity. While the red-shifted component seems missing in the observed spectrum.
%The tight upper limit of the red-shifted iron line intensity in the observed spectrum could be explained by either the intrinsic asymmetry of the hot wind or due to the line-of-sight absorption. 
As discussed in \citet{2021NatAs...5..928S}, the gas density of the wind above and below the equatorial plane can deviate from each other at a level of $\sim25$\% on the dynamical timescale ($\sim$10 yr) of the wind, which can lead to a $\sim50$\% difference between the observed redshifted and blueshifted components. 
While such an explanation was offered for the moderate inequality between the observed redshifted and blueshifted Fe XXVI Ly$\alpha$ lines in M81*, the substantially weaker redshifted component in the present case of NGC\,7213 disfavors an intrinsic variability as the primary cause.  
%However, the almost absence of the red-shifted component favors the absorption scenario.
%The hot wind itself is almost transparent to the highly ionized iron lines emitted from both sides. 
%Hot accretion flow inside the truncation radius is optically thin while the standard thin disk outside the truncation radius is optically thick. 

A second tunable parameter in the synthetic spectrum is the Fe abundance. Initially, we have assumed a solar abundance as adopted in the \emph{apec} models (Section~\ref{subsec:spec}).
However, the equivalent width of the Fe XXVI Ly$\alpha$ line derived from the synthetic spectrum is 32 eV, which is only about half of the observed value (61.2 eV). 
Hence we adopt a synthetic spectrum with the Fe abundance being twice solar, which leads to an equivalent width of 45 eV for Fe XXVI Ly$\alpha$ with the $C$-stat improved by ${\Delta}C =7$. 
We emphasize that the abundance cannot be strongly constrained, due to a degeneracy in the continuum between the wind  bremsstrahlung and contribution from the hot accretion flow (and/or a jet, if present) as represented by the power-law.
%We have further verified that the synthetic wind spectrum does not violate the observed spectrum when extrapolated to lower energies.  
%The twice solar abundance choice for Fe is also supported by the non-detection of fluorescent emission lines for neutral element like Si, S, Ar and strong Fe K$\alpha$ line in contrast.
%The 3-$\sigma$ upper limit of relative abundance  $\frac{Z_{\rm X}}{Z_{\rm Fe}}$ over Fe can be constrained to $<1.1$ for Si, $<0.6$ for S and $<0.6$ for Ar, based on the Eq (1) in \citet{2016MNRAS.455L..26X}.
%Thus we chose the Fe abundance to be twice of solar abundance that matches the observed spectrum best.

Given the best-matching synthetic wind spectrum, the outflow rate of the hot wind can be evaluated according to the simulated  density and radial velocity distributions.
We find an outflow rate of $\sim0.08\rm~M_{\odot}~yr^{-1}$.
Interestingly, this value is comparable to the inflow rate of $0.07-0.2\rm~M_{\odot}~yr^{-1}$ estimated for warm ionized gas within the central $\sim100$ pc traced by H$\alpha$ emission \citep{2014MNRAS.438.3322S}.
Theoretically, the hot wind is expected to carry away the bulk of the inflowing mass in the hot accretion flow \citep{2015ApJ...804..101Y}. The good agreement between the inferred outflow and inflow rates lends strong support to the wind scenario.
%It is consistent with the value derived from the {\it apec} model.

%On the other hand, the truncation radius is anti-correlated with the accretion rate \citep{2014ARA&A..52..529Y}. The Eddington ratio of NGC 7213 is of about one order of magnitude larger than M81$^*$, the putative truncation radius $r_{\rm tr,NGC7213}$ should be smaller than $r_{\rm tr,M81}$. The large inclination angle ($\sim80^{\circ}$) derived from the synthetic wind spectrum together with a smaller truncation radius makes a majority fraction of the wind in the redshifted side in NGC 7213 shaded by the disk along line-of-sight, especially regions between $10^{4}-10^{5}\rm~r_{g}$ where most of the Fe XXVI line emission originates, as shown in Figure \ref{fig:AD_absorb}. 
%In M81$^{*}$ case, the proper orientation enables the redshifted Fe XXVI photons to be visible with a small viewing angle ($\sim15^{\circ}$).

\section{Discussion} \label{sec:discuss}

\subsection{Alternative origins of the high-velocity hot plasma} \label{subsec:alter}
\subsubsection{Stellar activities} \label{subsec:stellar}
In principle, the high-velocity Fe lines may be produced by stellar activities such as young supernova remnants (SNRs) or massive star binaries with strong colliding winds.
%In extreme cases, an intense nuclear starburst may induce a multi-phase outflow at a velocity of $\sim10^3$ km/s \citep{2013MNRAS.429.3439E}.
However, the X-ray luminosity of the 16-keV plasma, $4\times10^{41}\rm~erg~s^{-1}$, is orders of magnitude higher than possible for a single young SNR \citep{2012MNRAS.419.1515D}. 
Theoretical investigation \citep{2016MNRAS.456.2537R} also shows that the cumulative X-ray luminosity from a collection of SNRs evolving within the sphere of influence of a $\sim$10$^8\rm~M_{\odot}$ SMBH ($\sim$100 pc in radius) does not exceed $\sim$10$^{38}\rm~erg~s^{-1}$. 
Similarly, colliding wind binaries cannot reach the observed X-ray luminosity and plasma temperature \citep{2012ASPC..465..301G}.

To further evaluate the collective X-ray emission from the circumnuclear stellar populations, we first estimate the stellar mass and star formation rate (SFR) in the HETG spectral extraction region.  
%The 2-10 keV X-ray luminosity of the 13-keV hot plasma derived in Section \ref{subsec:spec} has far exceeded the averaged energy output either from a single supernova event. Additionally, we estimated the birthrate of supernova in the nuclear region of NGC 7213. 
The stellar mass within a projected radius of 265 pc is found to be $\sim1\times10^{10}\rm~M_{\odot}$, based on the $V$-band surface brightness distribution of NGC\,7213 \citep{2005AJ....129.2138L} and assuming a $V$-band mass-to-light ratio of 5.4 typical of Sa galaxy \citep{1979ARA&A..17..135F}. 
%The corresponding birth rate of Type Ia supernova is  $$R_{\rm SN}^{\rm Ia}\sim3.5\times10^{-5}\rm~yr^{-1}$, according to the empirical relation of \citep{2005A&A...433..807M}. 
The SFR is estimated to be $\lesssim0.02\rm~M_{\odot}~yr^{-1}$, based on the measured luminosity of narrow H$\alpha$ line from the nucleus of NGC\,7213 \citep{1984ApJ...285..458F} and the empirical relation between SFR and H$\alpha$ luminosity \citep{1998ARA&A..36..189K}. 
%For a Chabrier/Kroupa initial mass function, the associated birth rate of core-collapse supernova is $R_{\rm SN}^{\rm cc}\lesssim2.5\times10^{-3}\rm~yr^{-1}$, assuming the average lifetime of the progenitor of core-collapse SN is $\sim2\times10^7\rm~yr$. 
%The possibility of multiple contemporaneous supernovae events that cause the strong hot plasma component could also be excluded.
%On the other hand, the colliding wind from stellar binary systems can hardly reach the temperature as high as $\sim10$ keV. 
%Stellar population in the nuclear region contributes to an 2-10 keV X-ray luminosity $L_{2-10}^{star}\sim8.3\times10^{37}\rm~erg~s^{-1}$ according to the empirical relation described in \citep{2021NatAs...5..928S} and reference therein. 
These estimates then provide an upper limit in the X-ray luminosity contributed by both old and young stellar populations, $L_{\rm 2-10~keV}^{\rm star} \lesssim 9.4\times10^{38}\rm~erg~s^{-1}$, according to the empirical relation given by \citet{2010ApJ...724..559L}: $L_{\rm 2-10~keV}^{\rm star} = \alpha M_* + \beta {\rm SFR}$, where $\alpha = (9.05\pm0.37)\times10^{28}{\rm~erg~s^{-1}~M_\odot^{-1}}$ and $\beta = (1.62\pm0.22)\times10^{39}{\rm~erg~s^{-1}~(M_\odot~yr^{-1})}^{-1}$. 
This is orders of magnitude lower than the observed X-ray luminosity of the 16-keV plasma. 
Moreover, the circumnuclear star formation is expected to produce a 0.5--2 keV luminosity of  $L_{0.5-2}^{\rm SFR} = 9\times10^{37}({\rm SFR}/0.02{\rm~M_{\odot}~yr^{-1}})\rm~erg~s^{-1}$, based on the empirical relation of \citet{2003A&A...399...39R}. This is about two orders of magnitude lower than the observed X-ray luminosity of the low-$T$ component (Section \ref{subsec:spec}).
Hence we conclude that stellar activities cannot account for the thermal components in the HETG spectrum.

\subsubsection{AGN Photoionization} \label{subsec:phoion} 
Alternatively, the highly ionized Fe lines may be produced by AGN photoionization.
We evaluate this possibility using the photoionization code {\it Cloudy} (ver c17.02; \citealp{2017RMxAA..53..385F}). The code calculates the ionization state of an isotropic and uniform gas cloud photoionized by a central source, which has an intrinsic X-ray spectrum, $\phi(\nu)$, same as the baseline power-law continuum (Section~\ref{subsec:cont}), i.e., with a photon-index of 1.723 and 0.5--10 keV luminosity of $2.4\times10^{42}\rm~erg~s^{-1}$. We note that the X-ray luminosity has varied by a factor of $\sim4$ over a period of $\sim$38 yr \citep{2018MNRAS.475.1190Y}, but our conclusion below is not affected by this moderate variability.
The inner radius of the cloud is set to be 0.001 pc ($\sim260~r_{\rm g}$), while the outer radius varies with different assumed values of column density $N_{\rm c} = n_{\rm c}(r_{\rm out}-r_{\rm in})$, where $n_{\rm c}$ is the cloud density assumed to be constant for simplicity. 
We note that adopting a radially-decreasing density distribution does not significantly affect our following conclusion.
The ionization parameter at the illuminated face of the cloud is expressed by $U_{\rm X}=\int^{\rm 10keV}_{\rm 2keV}\frac{\phi(\nu)d\nu}{4\pi r_{\rm in}^2cn_{\rm c}}$. 
%$\phi(\nu)$ is the photon flux received by the inner face of the cloud from central AGN. The incident total luminosity over 0.5 - 10 keV band was set to be $2.4\times10^{42}\rm~erg~s^{-1}$, equal to the observed unabsorbed X-ray luminosity.
The lack of reflection component in the broadband X-ray spectrum indicates a Compton-thin case. 
Hence we explore the plausible parameter space of $N_{\rm c} =10^{21}-10^{24}\rm~cm^{-2}$ and ionization parameters $U_{\rm X}
=1-7$. 
For a given set of $N_{\rm c}$ and $U_{\rm X}$,
the outward luminosity of a certain atomic line is then calculated using the theoretical line emissivity from ATOMDB and assuming an abundance of twice solar, to be consistent with the favored abundance in the synthetic wind spectrum.

\begin{figure*}[ht!]
\centering\includegraphics[width=0.49\textwidth]{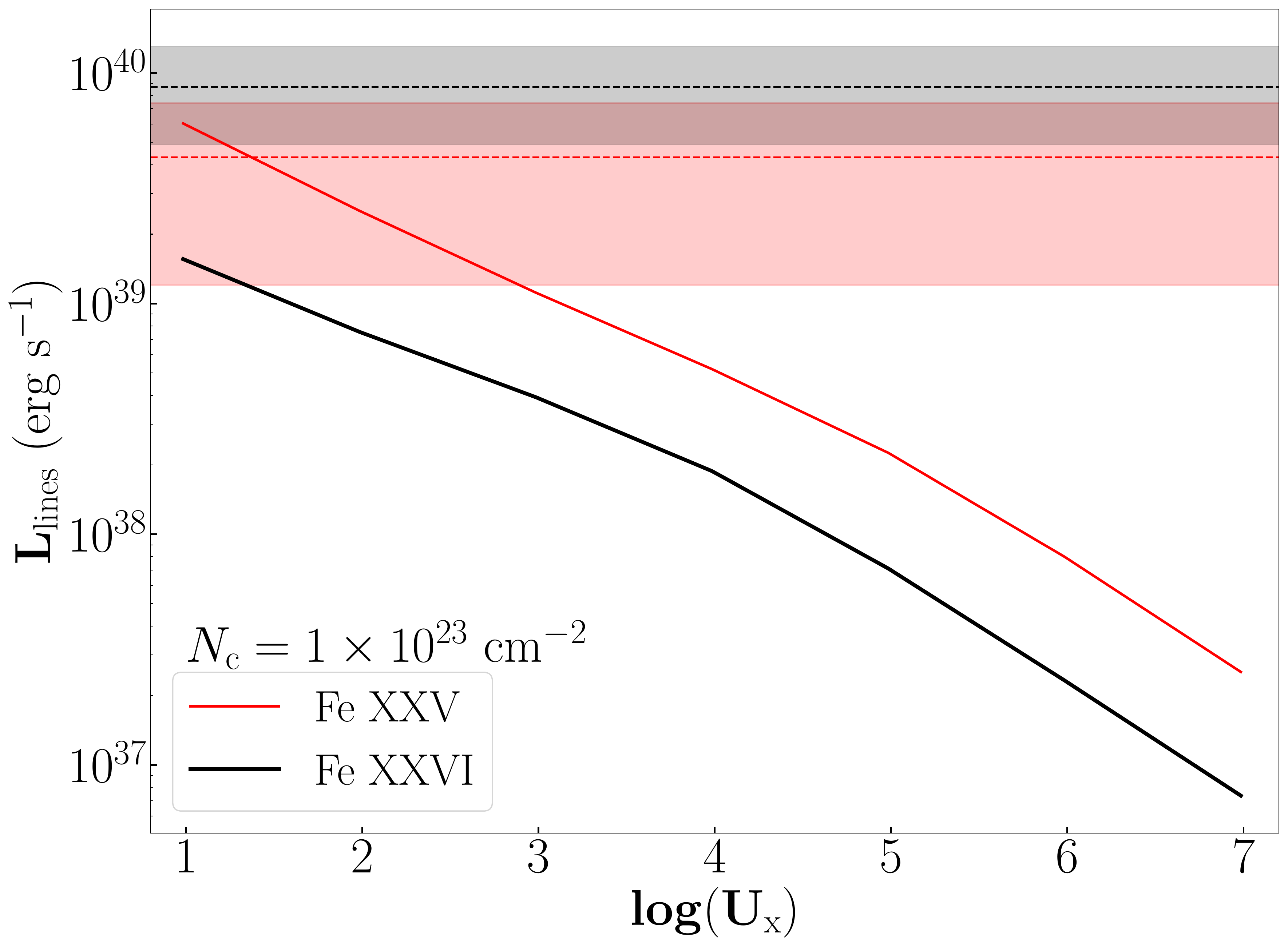}
\centering\includegraphics[width=0.49\textwidth]{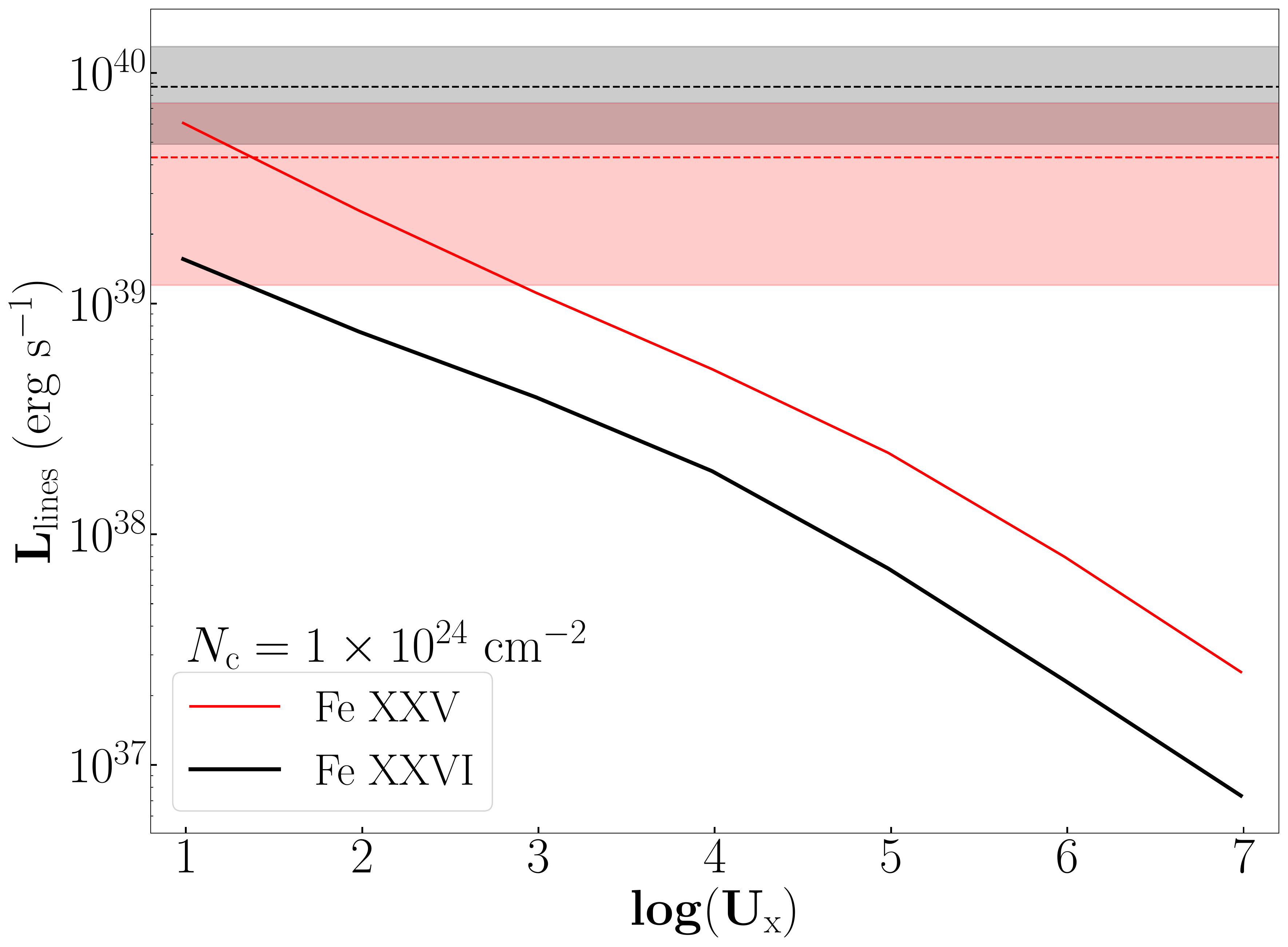}
\\
\centering\includegraphics[width=0.49\textwidth]{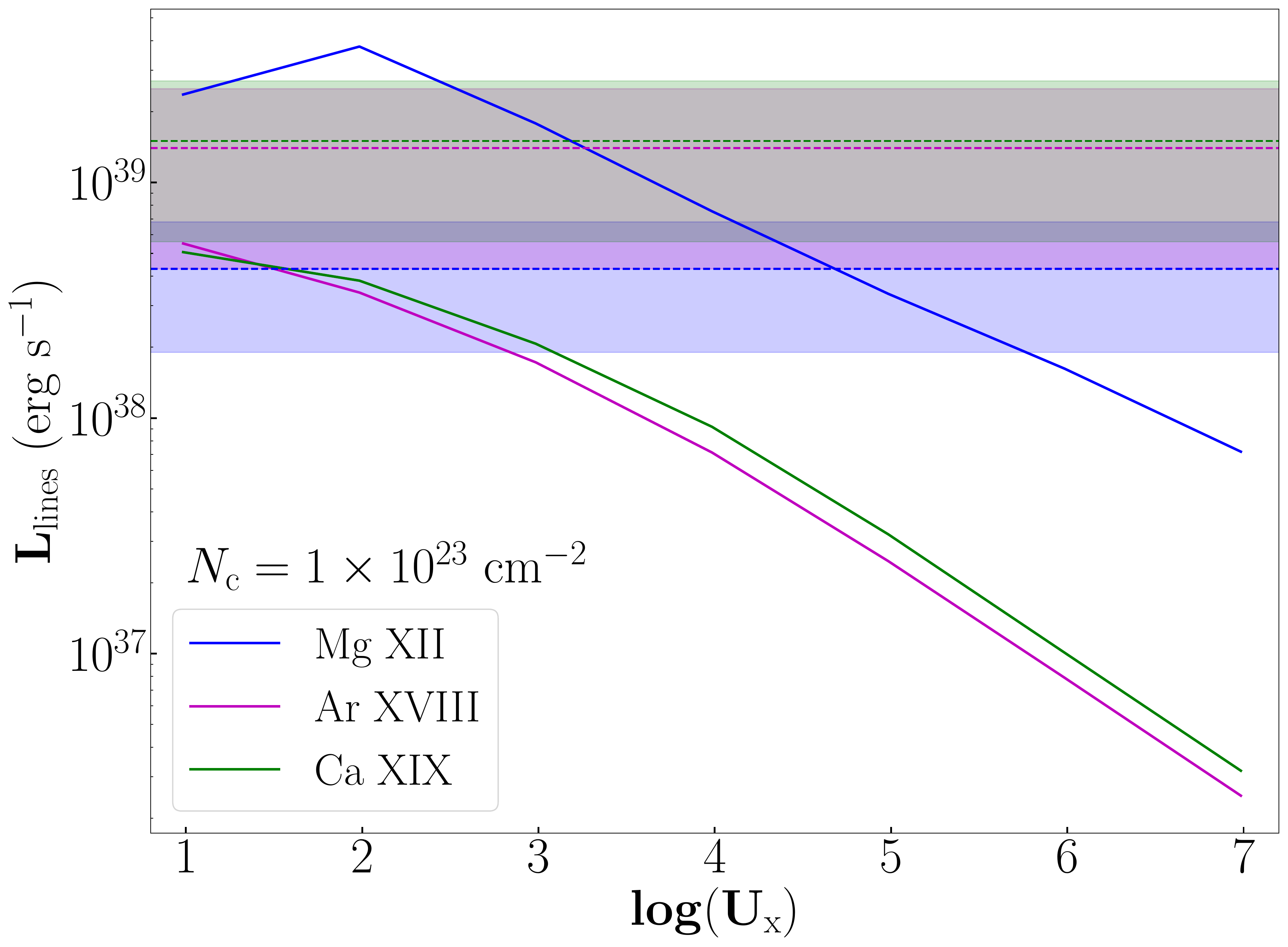}
\centering\includegraphics[width=0.49\textwidth]{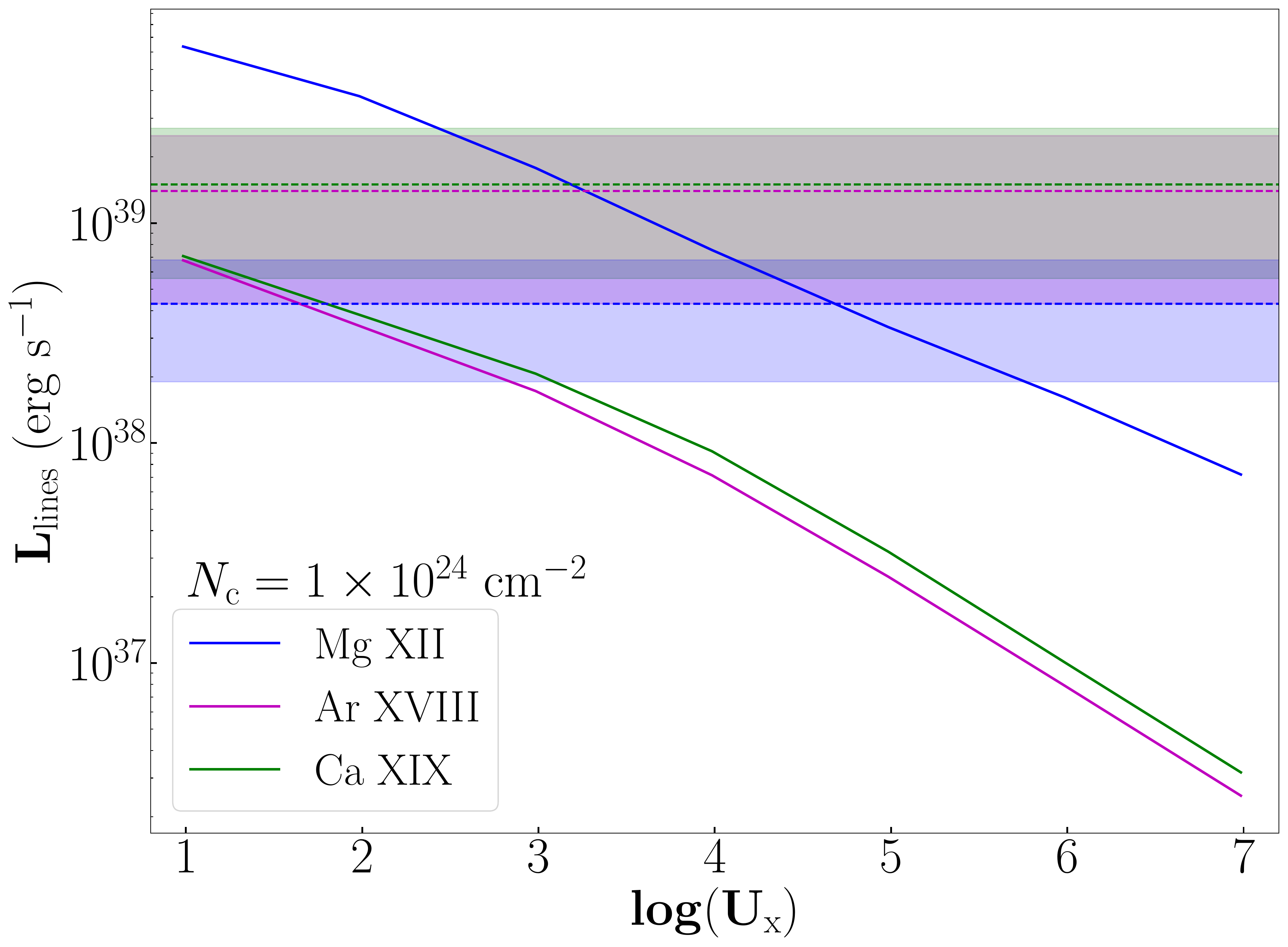}
\caption{The predicted line luminosity induced by AGN photoionization, as a function of the ionization parameter. The upper panels are for the Fe XXV K$\alpha$ and Fe XXVI Ly$\alpha$ lines, and the lower panels are for the Mg XII Ly$\alpha$, Ar XVIII Ly$\alpha$ and Ca XIX K$\alpha$ lines. The left and right panels assume a gas column density of $10^{23}\rm~cm^{-2}$ and $10^{24}\rm~cm^{-2}$, respectively.
The horizontal dotted lines and the adjacent stripes mark the observed luminosity and 90\% uncertainty.
\label{fig:phoion}}
\end{figure*}

The predicted line luminosities are shown in Figure~\ref{fig:phoion}. 
We find that the predicted Fe XXVI Ly$\alpha$ line from a photoionized cloud falls short by a factor of a few to a few hundred to the observed luminosity, for any reasonable combination of $N_{\rm c}$ and $U_{\rm X}$.
The predicted Fe XXV K$\alpha$ line is also typically weak and only comes closer to the observed luminosity at small values of $U_{\rm X}$.
Therefore, the hot plasma traced by the highly-ionized Fe lines are highly unlikely due to photoionization by the LLAGN. We note that \citet{2008MNRAS.389L..52B} also disfavored an origin of photoionization for the Fe lines based on the argument that the resonant transition dominates the Fe XXV triplet.
Line luminosities are also predicted for the three low-$Z$ lines detected: Mg XII Ly$\alpha$, Ar XVIII Ly$\alpha$ and Ca XIX K$\alpha$. 
Among them, the Mg XII Ly$\alpha$ line can have a predicted luminosity comparable to the observed value, whereas the other two predicted lines typically fall short to the observed value except with the lowest $U_{\rm X}$. This suggests that photoionization by the LLAGN can at best account for a fraction of the low-$Z$ lines.
%While the Ar..... lines indicated by plasma with lower temperature and maybe further to the central SMBH can be generated by photoionization of either AGN or stellar wind.

%{\bf Fig for Cloudy simulation. 4 panels (1e+21, 1e+22, 1e+23, 1e+24), each panel $L_x$-$\xi$, 5 lines}

%{\bf (Should we discuss the G ratio?)}

%We further did a rough estimate of the wind momentum flux $4\pi R^2<\rho><v>^2$ at a distance of $R$. The average mass density of the wind $<\rho>=\frac{<n>}{\mu_{\rm m}}$ has been evaluated from \emph{apec} model fitting (Table \ref{tab:mod_fit}). $\mu_m$ is the averaged atom mass. To determine the averaged hydrogen number density $<n>$ of the wind from \emph{apec} normalization, the volume that the wind occupies matters. Assuming its filling factor $\sim1$, the radius of the spherical wind could be constrained by line-of-sight column density.
%Ursini et. al. reported that the best-fit column density with MyTorus model of the putative Compton-thin torus surrounding NGC 7213 is $N_H\sim5\times10^{23}\rm~cm^{-2}$ using NuSTAR observation \citep{2015MNRAS.452.3266U}. 
%Thus the wind momentum flux is $\sim3.7\times10^{33}\rm~g~cm~s^{-2}$, which is about one order of magnitude larger than the photon momentum flux $\frac{L_{bol}}{c}\sim(3-6)\times10^{32}\rm~g~cm~s^{-2}$ of NGC 7213.

\subsubsection{The hot accretion flow} \label{subsec:inflow}
In principle, the hot accretion inflow inside the truncation radius can also produce highly-ionized Fe lines with a substantial Doppler shift. 
To test such a possibility, we utilize a set of 3D general relativity (GR) MHD simulation of the hot accretion flow with Kerr metric performed with the Atheta++ code \citep{2016ApJS..225...22W}. 
The simulation is scale-free in black hole mass, hence we can scale the simulation data to $M_{\rm BH}=10^8\rm~M_{\odot}$. Although the 3D GRMHD simulation only extends from the event horizon to 100 $r_{\rm g}$, all physical quantities of the hot accretion flow, such as density, temperature and velocity have a tight scaling relation in the radial direction, which allow us to extrapolate these quantities from 100 $r_{\rm g}$ to 2000 $r_{\rm g}$. 
An inflow rate of $0.08\rm~M_{\odot}~yr^{-1}$ at $2000~r_{\rm g}$ is adopted to normalize the gas density, which is consistent with the derived outflow rate in the hot wind (Section~\ref{subsec:compare}).
%which allow us to calculate the thermal X-ray emission, in particular the Fe lines, from the hot accretion flow.

We then follow the same procedure for the wind simulation to produce synthetic spectrum of the hot accretion flow for a given line-of-sight. 
The Fe abundance is set to be twice solar.
%{\bf We chose the same viewing angle of ($\sim80^{\circ}$) and iron abundance as the hot wind.} 
%{\bf The reason to exclude the inflow origin of the blueshifted iron lines is similar to that discussed in \citep{2021NatAs...5..928S}.}
We find that within $1000~r_{\rm g}$ the plasma temperature reaches  $>8\times10^{9}\rm~K$, which is so high that the gas becomes fully ionized; the bulk of Fe XXV and Fe XXVI ions exist in the radial range of $1000-2000~r_{\rm g}$, where the temperature drops to $\sim10^9$ K.
However, the predicted equivalent width of the Fe XXVI Ly$\alpha$ line is only $\sim15\rm~eV$, much low than the observed value of 61.2 eV. 
Moreover, 
due to the wide spread of the projected velocity in this radial region, the resultant line profile takes a broad and flat-topped shape, which is inconsistent with the observed narrow line profile. 
We also find it difficult to produce a large flux ratio as observed between the blueshifted and redshifted components, since the inflow itself is optically-thin and is barely obscured by the truncated disk.
These discrepancies between the synthetic and observed spectra persist for a number of viewing angles tested.
Hence we conclude that the blueshifted highly-ionized Fe lines are not originated from the hot accretion flow.

%\textcolor{blue}{We took the assumption that the outflow rate at the innermost radius of the hot wind should be equal to the inflow rate at the same radius of the hot accretion inflow. We found the ...}

\subsection{Comparison with M81* and implications for LLAGN feedback} \label{subsec:compare}
In the previous sections, we have shown that the blueshifted lines from highly ionized Fe are best understood as arising from a hot wind lunched from the hot accretion flow onto the SMBH in NGC\,7213.
A mass outflow rate of $\sim0.08\rm~M_{\odot}~yr^{-1}$ is inferred for this hot wind (Section~\ref{subsec:compare}).
We further estimate an associated kinetic power of $\sim3\times10^{42}\rm~erg~s^{-1}$. 
This is equivalent to $\sim15$\% of the SMBH's bolometric luminosity \citep{2018MNRAS.475.1190Y}, a fraction very similar to that found for the hot wind in M81* \citep{2021NatAs...5..928S}.
The momentum flux of the hot wind is estimated to be $\sim4\times10^{33}\rm~g~cm~s^{-1}$, which is $\sim$6 times the photon momentum flux from the LLAGN.
This is again similar to the case of M81*.
In view that M81* and NGC\,7213 encompass a substantial range in the Eddington ratio ($3\times10^{-5} -10^{-3}$), this similarity may suggest a quasi-universal mechanical output of the hot wind with respect to the SMBH's radiative output, which deserves further examination when detection of the hot wind in more LLAGNs becomes available. 
In any case, the kinetic power and momentum flux of the hot wind can provide an efficient and long-lasting means of feedback from weakly accreting SMBHs that are prevalent in normal galaxies. 

Among the crucial open questions about LLAGN feedback is whether the hot wind can deposit its kinetic energy and momentum into a substantial volume of the host galaxy, as required to effectively quench star formation in cosmological simulations  \citep{2017MNRAS.465.3291W,2018MNRAS.473.4077P}.
Numerical simulations predict that the hot wind will interact with the circumnuclear medium, releasing its kinetic energy and momentum and potentially creating strong shocks \citep{2019ApJ...885...16Y}.
The 0.8-keV plasma found in the HETG spectrum might be evidence for such an interaction, which is likely the manifestation of a circumnuclear hot gas shock-heated by the outward propagating hot wind.
We note that circumnuclear hot gas with a similar temperature is also evident in the case of M81* \citep{2021NatAs...5..928S}.
%$2.15\times10^{22}\times\sqrt{V}$
The internal energy of this circumnuclear hot gas in NGC\,7213 is $\lesssim1\times10^{54}\rm~erg$, assuming that it fills a spherical volume with a radius of 265 pc (i.e., the HETG spectral extraction region). This translates to a reasonably short time of $\sim1\times10^{4}\rm~yr$ if the entire kinetic power of the hot wind were converted into this internal energy.

%and it is consistent with the dynamical time-scale of the wind. 

The hot wind could have propagated to larger distances from the LLAGN. 
The zeroth-order spectrum of the off-nucleus region examined in Section~\ref{subsec:circum_reg} allows for the presence of a 0.8-keV plasma out to $\sim550$ pc, but with a substantially lower surface brightness. 
%contributing to the 1.2-keV component in the off nucleus spectrum in Section \ref{subsec:circum_reg} about $\sim250$ pc away from central black hole.
This may be understood as either the hot wind having been impeded and had the bulk of its kinetic power exhausted by the circumnuclear medium, or a paucity of gas in the off-nucleus region. 
Interestingly, recent ALMA CO observations reveal the presence of non-rotating molecular gas features at distances of few hundred pc from the nucleus \citep{2020A&A...641A.151S}, which were interpreted as a molecular gas outflow with a velocity of $\lesssim100$ km~s$^{-1}$ and a mass outflow rate of $0.03\pm0.02\rm~M_\odot~yr^{-1}$. 
While \citet{2020A&A...641A.151S} favored stellar activity as the mechanism responsible for the outflow, 
the possibility that this molecular gas is accelerated and entrained by the hot wind deserves further investigation.  

At present, direct evidence for the hot wind from a weakly accreting SMBH is only found in M81* and NGC\,7213. This may be attributed to two limiting factors. On the one hand, high-resolution X-ray spectra are currently only available for a handful of nearby LLAGNs (Appendix~\ref{sec:sample}). Among these sources, M81* and NGC\,7213 stand out with better data quality to allow for the detection of highly-ionized Fe lines. 
On the other hand, even with a deep X-ray spectrum, the presence of a strong X-ray continuum from  a relativistic jet (e.g., in the case of Cen A and M87) may easily mask the emission lines.
Future high-resolution X-ray spectroscopic observations by {\it Chandra}, as well as planned missions such as XRISM and Athena, hold promise to confirm the universality of a hot wind from LLAGNs.

%Molecular outflows with velocity of $\sim100\rm~km~s^{-1}$ about $\sim150$ pc away from the NGC 7213 nucleus along the minor axis has been detected by ALMA CO(2-1) observation \citep{2020A&A...641A.151S}. Although the mass outflow rate $\dot{m}_{\rm mol}\sim0.03-0.05\rm~M_{\odot}~yr^{}-1$ is consistent with our estimation from the hot wind,
%Salvestrini et. al. claim that stellar-like driven mechanism dominants in the molecular outflows together with very little contribution from central AGN. 
%Numerical simulations has also predicted that hot wind can hardly reach beyond 10 pc. By depleting gas in the inner region and suppress gas outward, it can enhance star formation in the surrounding region.
%Lacking direct evidence, we would not suggest a strong connection between the hot wind and the molecular outflows in NGC 7213.

%Hot wind detected in NGC 7213 has been on average two orders of magnitude more powerful than that in M81$^{*}$.
%With an Eddington ratio close to the threshold of 0.01, NGC 7213$^*$ could provide a novel view of AGN feedback as a connection between the nearly quiescent nuclei (like Sgr A$^*$, M81$^*$) and nuclei with moderate accretion rate.

\section{Summary}
\label{sec:sum}
We have searched the {\it Chandra} archive for ACIS/HETG observations of LLAGNs that may reveal potential evidence for a putative hot wind driven by a hot accretion flow and characterized by highly-ionized iron lines, as recently established in the case of M81*.  
Only one source, the LLAGN in NGC\,7213, manifests itself with X-ray spectral features similar to those found in M81*.
We have performed an in-depth X-ray spectral analysis for NGC\,7213, as well as MHD simulations of the hot wind tailored to the empirical conditions of the weakly accreting SMBH in this galaxy. 
Our main findings are as follows.

\begin{itemize}
    \item[(i)] A blind search results in the detection of a number of emission lines from the {\it Chandra} 1st-order HEG/MEG spectra. In particular, Fe XXVI Ly$\alpha$ and Fe XXV Ka$\alpha$ lines with a blue-shifted velocity of $\sim1100\rm~km~s^{-1}$ are identified. The flux ratio between the two lines suggest that they are tracing a hot plasma with a characteristic temperature of $16\pm6$ keV. The spectra show no significant redshifted counterpart to this plasma. In addition, several emission lines from low-$Z$ elements indicate the presence of a $\sim$0.8 keV plasma, which may be tracing a circumnuclear hot gas.
    \item[(ii)] The synthetic X-ray spectrum of the hot wind from the MHD simulation with a viewing angle of $\sim80^{\circ}$ and an Fe abundance twice solar can well match the blueshifted, highly-ionized Fe lines. In this case, the unseen redshifted counterpart is most likely due to obscuration by the truncated disk along the line-of-sight. All plausible alternative origins considered for the Fe lines, such as stellar activity, AGN photoionization and the hot accretion flow itself, are quantitatively disfavored. This makes a strong case that a hot wind exists in the  nucleus of NGC\,7213, presumably driven by the hot accretion flow.
    \item[(iii)] The mass outflow rate of the hot wind is estimated to be $\sim0.08\rm~M_{\odot}~yr^{-1}$, quite comparable to the inflow rate of ionized gas estimated for warm ionized gas within the central $\sim$100 pc traced by H$\alpha$ emission. This agreement between the outflow and inflow rates lends further support to the hot wind scenario for NGC\,7213.
     \item[(iv)] The wind carries a kinetic energy of $\sim3\times10^{42}\rm~erg~s^{-1}$ and a momentum flux of $\sim4\times10^{33}\rm~g~cm~s^{-1}$, thus providing an important means of kinetic feedback to the environment. 
\end{itemize}

Future X-ray and optical spectroscopic observations in a spatially-resolved fashion will be crucial to determining to what extent and at what efficiency the hot wind can affect the interstellar medium of the host galaxy. 

%We have almost pushed to the Chandra HETG detection limit of hot wind in LLAGN. Due to the low-luminosity nature of these potential candidates, high spectral and spatial resolution together with large effective area over 5-8 keV band would be required. Upon the launch of next-generation X-ray IFU telescope like Athena or Lynx X-ray observatory, we would expect an expansion in the sample of hot wind embedded LLAGN. 

\begin{acknowledgments}
This research made use of observations taken from the {\it Chandra} X-ray Observatory, software package CIAO provided by the {\it Chandra} X-ray Center and spectral fitting package XSPEC,
and the High Performance Computing Resource in the Core Facility for Advanced Research Computing at Shanghai Astronomical Observatory. 
Z.L. and F.S. acknowledges support by the National Key Research and Development Program of China (2017YFA0402703) and Natural Science Foundation of China (grant 11873028).
F.Y. and B.Z. are supported in part by the Natural Science Foundation of China (grants 11633006, 12133008, 12192220, and 12192223). 
\end{acknowledgments}

\software{CIAO \citep{2006SPIE.6270E..1VF}, XSPEC \citep{1996ASPC..101...17A}, NuSTARDAS (v1.9.2), ZEUS-MP/2 (v1.5.19; \citealp{2006ApJS..165..188H}), Cloudy (vc17.02; \citealp{2017RMxAA..53..385F}), Atheta++ code \citep{2016ApJS..225...22W}}

\appendix
\section{sample selection} \label{sec:sample}
We examined publicly available {\it Chandra} ACIS/HETG observations of nearby LLAGNs for potential signatures for the putative hot wind.
We found a total of 79 well-classified AGNs with existing ACIS/HETG observations.
For these AGNs, we calculated the Eddington ratio, $\lambda_{\rm Edd}=L_{\rm bol}/L_{\rm Edd}$, where 
$L_{\rm Edd}=1.26\times10^{38}(M_{\rm BH}/M_{\rm \odot})\rm~erg~s^{-1}$ is the Eddington luminosity. 
%We evaluate the Eddington luminosity 
For sources without $M_{\rm BH}$ measurement in the literature, we assumed $M_{\rm BH}\simeq10^7\rm~M_{\odot}$.
We also assumed that the 2--10 keV X-ray luminosity, looked up from the literature, accounts for $\sim10\%$ of the bolometric luminosity $L_{\rm bol}$. 
%It is inevitable that the unabsorbed 2-10 keV flux of some sources is subjected to uncertain intrinsic absorption. 

%for LLAGN and gradually increases to $\sim50\%$ for AGN with $L_{\rm bol}\sim10^{46}\rm~erg~s^{-1}$\citep{2012MNRAS.425..623L}. Since the maximum 2-10 keV X-ray luminosity in our sample dose not exceed the $10^{46}\rm~erg~s^{-1}$, we gave the upper limit estimation for both $\lambda_{\rm Edd}$ by assuming the bolometric luminosity is 10 times the unabsorbed $L_{\rm 2-10~keV}$.

\begin{figure*}[ht!]
\centering\includegraphics[width=0.95\textwidth]{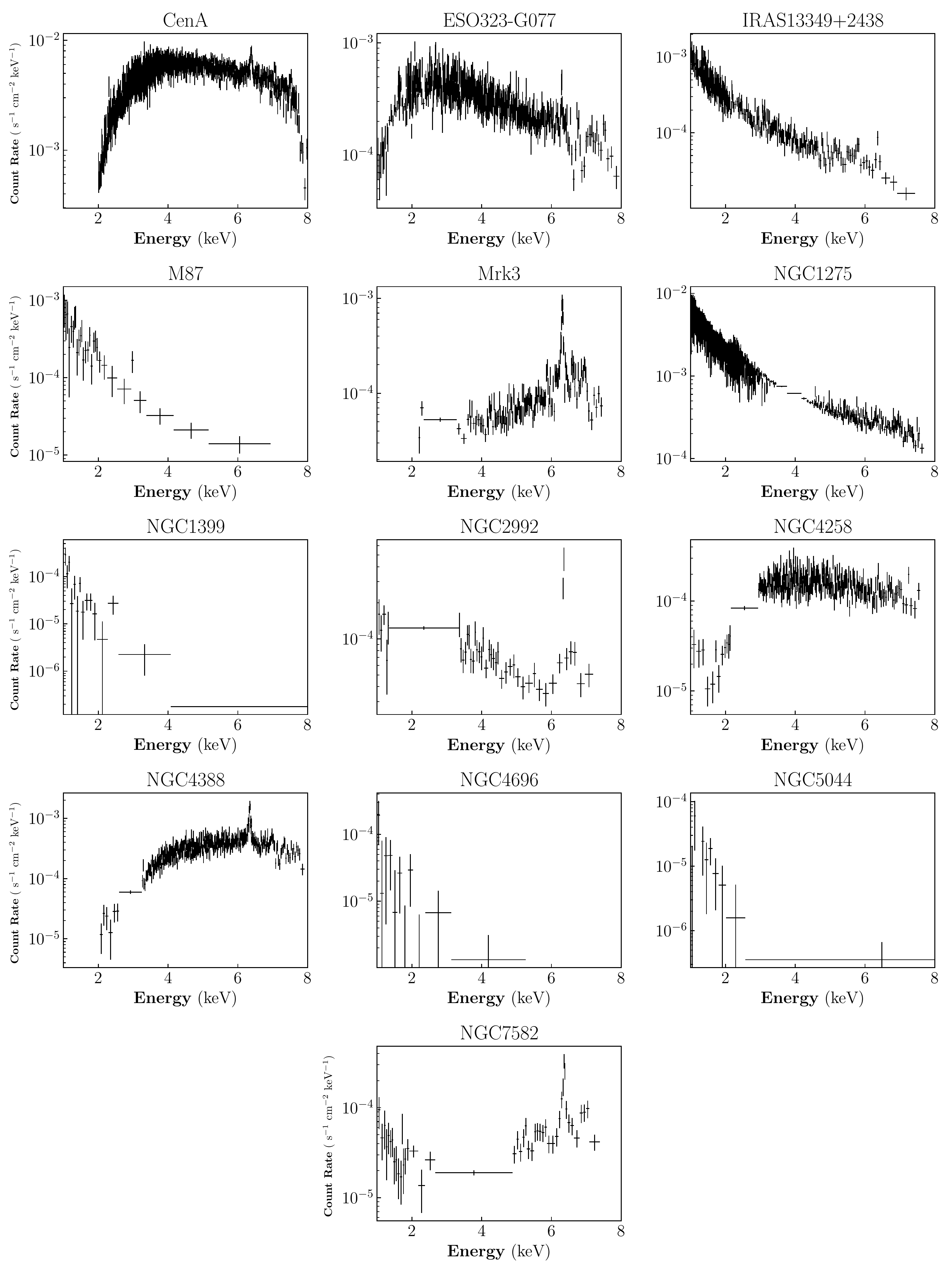}
\caption{The HEG spectra of 13 LLAGNs selected from the {\it Chandra} archive. The spectra are binned to have a signal-to-noise ratio great than 3 for better illustration.
The target name is provided in each panel.
\label{fig:sample}}
\end{figure*}

Among the 79 AGNs with ACIS/HETG observations, we classified 14 LLAGN candidates, which have $\lambda_{\rm Edd} \lesssim10^{-3}$. 
%Some physical properties of these selected LLAGNs are listed in \textbf{Table X (Table of physical properties of LLAGN sample needed or not?)}.
We have downloaded and reprocessed the ACIS/HETG data of these 14 targets following the same procedures described in Section~\ref{sec:data}. 
Unfortunately, only one target, NGC\,7213, exhibits clear evidence of emission lines from the H-like and He-like Fe, which are taken as a defining signature of the putative hot wind. 
The remaining 13 LLAGNs, including famous targets such as Cen A and M87, show no significant lines of highly-ionized Fe. 
For reference, the HEG spectra of these 13 LLAGNs are  shown in Figure \ref{fig:sample}.
%We used CIAO v4.13 and calibration files (v4.9.5) to reprocess the high-energy grating (HEG) and extracted the $\rm\pm1$-order spectra with default source and background region by \emph{tgextract}. Multi epoch spectra for a single source have been co-added and regroup to signal-to-noise ratio equals 3.

%\textbf{(Need some description for spectra for other LLAGN?)}
%Only NGC 7213 exhibits emission features around 6.5-7 keV, with a simple power-law continuum very similar to M81*. 
%So we did a detailed analysis on X-ray spectrum of NGC 7213 in the following sections. 
\section{Self-absorption of the hot wind to the Fe lines} \label{sec:depth}

To verify that self-absorption of Fe XXV K$\alpha$ and Fe XXVI Ly$\alpha$ lines in the hot wind can be neglected, we evaluate the optical depth \citep{1979rpa..book.....R}, $\tau(E)=N_{\rm i}\frac{\pi e^2}{m_{\rm e}c}f_{\rm lu}\phi(E)$, along a typical line-of-sight, where $N_{\rm i}$ refers to the column density of a certain ion (here Fe XXV or Fe XXVI), $f_{\rm lu}$ is the oscillator strength of electron transition from lower level $l$ to upper level $u$, $m_{\rm e}$ is the electron mass and $c$ is the speed of light.
%Photons emitted by transition $u-l$ of certain ions in the inner radius would fail to be absorbed by the same ions at outer radius, due to the bulk motion velocity of gas changes at different radius which shifts the peak energy of normalized 
The value of $N_{\rm i}$ as a function of temperature is obtained from ATOMDB.
The iron abundance is chosen to be twice solar, along with an average hydrogen column density of $\sim6\times10^{22}\rm~cm^{-2}$ from the wind simulation. 
The Voigt function $\phi(E)$ is numerically solved for the projected gas velocity along the line-of-sight. 

\begin{figure}[ht!]
\centering\includegraphics[width=0.6\textwidth]{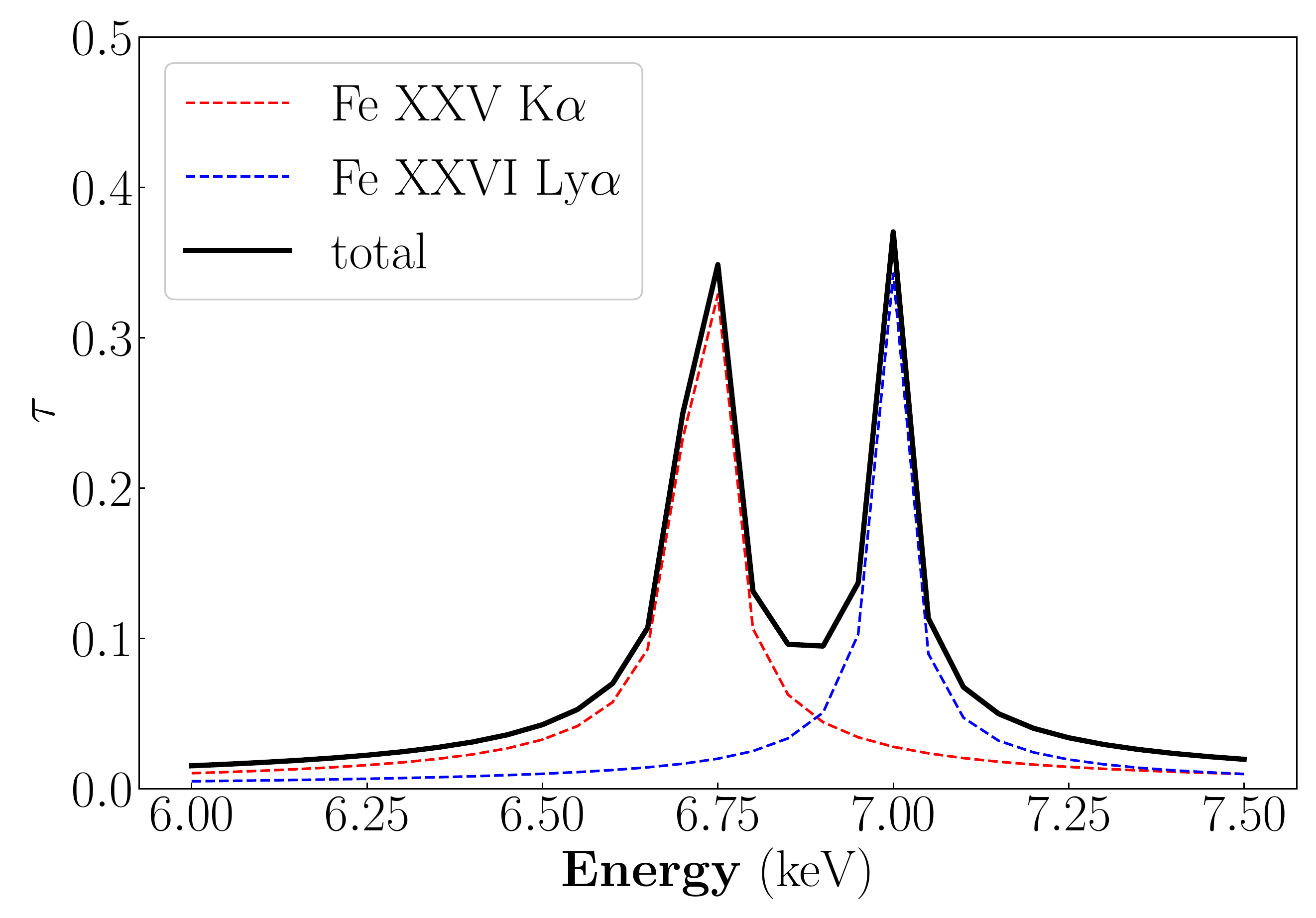}
\caption{The optical depth of the hot wind as a function of photon energy between 6--7.5 keV (black curve), for a viewing angle of 80$^{\circ}$. The two peaks are due to Fe XXV K$\alpha$ and Fe XXVI Ly$\alpha$. Self-absorption is only moderate in both lines.}
\label{fig:op_depth}
\end{figure}

The resultant optical depth of the hot wind for photon energies between 6--7.5 keV is shown in Figure \ref{fig:op_depth}. 
It can be seen that in the vicinity of the Fe XXV K$\alpha$ and Fe XXVI Ly$\alpha$ lines the optical depth is typically $\lesssim0.1$ and always $\lesssim$0.37.
We conclude that our assumption of negligible self-absorption in the hot wind is adequate.

%Table of HEG/MEG mutual blind-line search results [z-corrected] [upper MEG/ lower HEG]

%%[Chandra only] Table of observation log v1
%\begin{deluxetable*}{ccc}
%\tablenum{4}
%\tablecaption{Log of Chandra X-ray observations\label{tab:log_obs}}
%\tablewidth{0pt}
%\tablehead{
%\colhead{ObsID} & \colhead{Start Time} & \colhead{Exposure}\\
%\colhead{ } & \colhead{(UT)} & \colhead{(ks)}
%}
%%\decimalcolnumbers
%\startdata
%7742 & 2007-08-06 19:30:31 & 113.4\\
%8590 & 2007-08-09 12:25:28 & 34.56\\
%\enddata
%% already correlated for the redshift of NGC7213
%%\tablecomments{ }
%\end{deluxetable*}

%%%%%%Figures%%%%%%%%%%%%

%\clearpage
%\begin{figure*}[ht!]
%\plotone{.pdf}
%\caption{\textcolor{blue}{}
%\label{fig:inflow}}
%\end{figure*}


\bibliography{sample631}{}
\bibliographystyle{aasjournal}



\end{document}